\documentclass[sigconf]{acmart}
\usepackage{xcolor}
\definecolor{mouse}{HTML}{fbb4ae} 
\definecolor{keyboard}{HTML}{ccebc5} 
\definecolor{voice}{HTML}{decbe4} 
\definecolor{highlight}{HTML}{fed9a6} 
\definecolor{radial}{HTML}{b3cde3} 
\definecolor{medium}{gray}{0.90}
\definecolor{difficult}{gray}{0.70}
\newcommand{\sysName}[0]{Diana\xspace}
\newcommand{\DGOne}[0]{\textit{DG1 (familiar)}\xspace}
\newcommand{\DGTwo}[0]{\textit{DG2 (multimodal)}\xspace}
\newcommand{\DGThree}[0]{\textit{DG3 (human-like)}\xspace}
\newcommand{\DGFour}[0]{\textit{DG4 (orientation)}\xspace}
\newcommand{\DGFive}[0]{\textit{DG5 (guidance \& autonomy)}\xspace}
\newcommand{\DianaParticipant}[1]{D#1\xspace}
\newcommand{\BaselineParticipant}[1]{B#1\xspace}

\copyrightyear{2026}
\acmYear{2026}
\setcopyright{cc}
\setcctype{by}
\acmConference[CHI '26]{Proceedings of the 2026 CHI Conference on Human Factors in Computing Systems}{April 13--17, 2026}{Barcelona, Spain}
\acmBooktitle{Proceedings of the 2026 CHI Conference on Human Factors in Computing Systems (CHI '26), April 13--17, 2026, Barcelona, Spain}
\acmPrice{}
\acmDOI{10.1145/3772318.3791766}
\acmISBN{979-8-4007-2278-3/2026/04}
\graphicspath{{figures/}{pictures/}{images/}{./}}

\usepackage{enumitem}

\usepackage{svg}

\usepackage[fixed]{fontawesome5}

\usepackage{ellipsis}

\PassOptionsToPackage{hyphens}{url}

\usepackage{tabularx}
\usepackage{multirow}

\usepackage{xspace}

\usepackage{color, colortbl}

\usepackage{subfig}

\usepackage{soul}

\definecolor{visblue}{RGB}{132,166,178}
\definecolor{dashpurple}{RGB}{186,142,172}

\newif\ifshowchanges\showchangesfalse
\newcommand{\inserted}[1]{%
    \ifshowchanges
        \textcolor{blue}{{}#1}%
    \else
        #1%
    \fi}
\newcommand{\deleted}[1]{%
    \ifshowchanges
        {\textcolor{red}{\st{#1}}\normalfont}%
    \else\fi}

\begin{document}

%% ---------------------------------------------------------------------
%% Paper title and Authors
%% ---------------------------------------------------------------------

%% Title
\title[``Hey Dashboard!'': Supporting Voice, Text, and Pointing Modalities in Dashboard Onboarding]{``Hey Dashboard!'': Supporting Voice, Text, and Pointing Modalities in Dashboard Onboarding using Large Language Models}

%% Authors
\author{Vaishali Dhanoa}
\email{dhanoa@cs.au.dk}
\orcid{0000-0002-0493-8616}
\affiliation{%
  \institution{Aarhus University}
  \city{Aarhus}
  \country{Denmark}
}
\affiliation{%
  \institution{TU Wien}
  \city{Vienna}
  \country{Austria}
}

\author{Gabriela Molina Le\'{o}n}
\email{leon@cs.au.dk}
\affiliation{%
  \institution{Aarhus University}
  \city{Aarhus}
  \country{Denmark}
}

\author{Eve Hoggan}
\email{eve.hoggan@cs.au.dk}
\affiliation{%
  \institution{Aarhus University}
  \city{Aarhus}
  \country{Denmark}
}

\author{Eduard Gr\"{o}ller}
\email{groeller@cg.tuwien.ac.at}
\affiliation{%
 \institution{TU Wien \& VRVIS}
 \city{Vienna}
 \country{Austria}}

\author{Marc Streit}
\email{marc.streit@jku.at}
\affiliation{%
  \institution{Johannes Kepler University Linz}
  \city{Linz}
  \country{Austria}}

\author{Niklas Elmqvist}
\email{elm@cs.au.dk}
\affiliation{%
  \institution{Aarhus University}
  \city{Aarhus}
  \country{Denmark}
}

\renewcommand{\shortauthors}{Dhanoa et al.}

%% ---------------------------------------------------------------------
%% Abstract
%% ---------------------------------------------------------------------
\begin{abstract}
    Visualization dashboards are regularly used for data exploration and analysis, but their complex interactions and interlinked views often require time-consuming onboarding sessions from dashboard authors.
    Preparing these onboarding materials is labor-intensive and requires manual updates when dashboards change.
    Recent advances in multimodal interaction powered by large language models (LLMs) provide ways to support self-guided onboarding.
    We present \textsc{Diana} (Dashboard Interactive Assistant for Navigation and Analysis), a multimodal dashboard assistant that helps users for navigation and guided analysis through chat, audio, and mouse-based interactions.
    Users can choose any interaction modality or a combination of them to onboard themselves on the dashboard.
    Each modality highlights relevant dashboard features to support user orientation.
    Unlike typical LLM systems that rely solely on text-based chat, \textsc{Diana} combines multiple modalities to provide explanations directly in the dashboard interface.
    We conducted a comparative qualitative user study to understand the use of different modalities for different types of onboarding tasks and their complexities.
\end{abstract}



%% Keywords

%% ---------------------------------------------------------------------
%% Teaser
%% ---------------------------------------------------------------------

\begin{teaserfigure}
  \includegraphics[width=\textwidth]{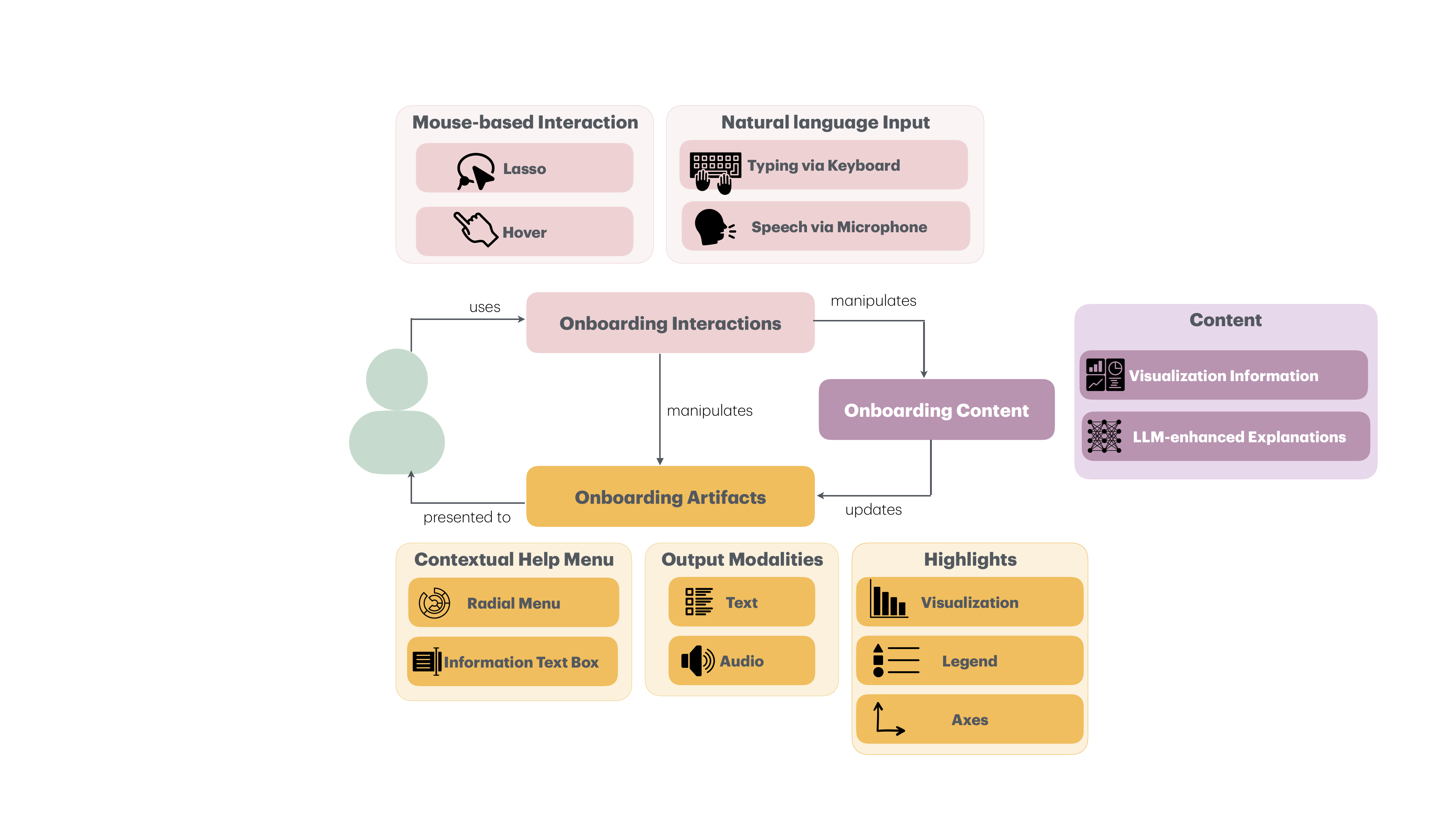}
  \caption{\textbf{Multimodal interaction for dashboard onboarding using LLMs.} \textnormal{Building on the process model for dashboard onboarding proposed by Dhanoa et al.~\cite{v_process_2022}, we extend the model with multimodal interactions and LLM-powered features for user onboarding.}}
  \Description{The figure shows how onboarding interactions, interface artifacts, and onboarding content relate within an AI-enhanced onboarding system. It shows users initiating actions such as hovering, clicking, typing, or speaking, which in turn manipulate the underlying dashboard content and activate artifacts like contextual help menus or radial menus. These artifacts are presented to the user as text, audio, and explanations generated by a language mode and this content updates based on additional user manipulation. The figure conveys a pattern of continuous feedback loops rather than a linear flow: user interactions trigger system responses, and those responses shape the next user actions, illustrating an integrated ecosystem of input, interpretation, and adaptive guidance.}
  \label{fig:teaser}
\end{teaserfigure}

\maketitle

%% ---------------------------------------------------------------------
%% Packages
%% ---------------------------------------------------------------------

%% ------medium---------------------------------------------------------------
%% Content
%% ---------------------------------------------------------------------
\section{Introduction}

Visualization dashboards have become essential tools for data exploration and decision-making across organizations~\cite{DBLP:journals/tvcg/SarikayaCBTF19}.
Modern dashboards integrate multiple linked views, complex filtering mechanisms, and sophisticated interaction techniques to support analytical workflows~\cite{bach23dashboardpatterns_tvcg}.
However, this complexity creates a significant barrier: users often struggle to understand how to navigate and effectively use these dashboards without guidance~\cite{walchshofer_transitioning_2023, ceneda_characterizing_2017, collins_guidance_2018}.
Dashboard authors accordingly invest substantial time creating \textit{onboarding} materials and conducting training sessions~\cite{v_process_2022, c_visualization_2019}, yet these resources quickly become outdated as dashboards evolve~\cite{DBLP:journals/tvcg/DhanoaHFEGS25}.
Existing automated onboarding solutions often fail to match the effectiveness of human experts who can provide contextual explanations, adapt to user questions, and demonstrate features using the actual data and visualizations present in the dashboard.

We present \sysName (Dashboard Interactive Assistant for Navigation and Analysis), a multimodal dashboard assistant that replicates the guidance provided by human onboarding experts.
Unlike text-only chatbots or onboarding tutorials (static or dynamic~\cite{DBLP:journals/tvcg/DhanoaHFEGS25}), \sysName combines multiple interaction modalities---voice input and output, mouse, and keyboard-based interaction---to provide \textit{situated explanations}~\cite{DBLP:journals/ivs/ChunduryYCMSE23} directly within the dashboard interface.
The system is based on the process model for dashboard onboarding described by Dhanoa et al.~\cite{v_process_2022} and uses LLMs for the unique case of onboarding (Figure~\ref{fig:teaser}). 
It employs a contextual radial menu that organizes dashboard features into logical structures, supports pointing gestures and lasso selection for highlighting specific charts or data points, includes a text-based natural language interface, and integrates speech recognition and synthesis for natural conversation.
\sysName delivers contextual help using the actual tasks, visualizations, and datasets from each specific dashboard, enabling users to learn through direct manipulation and exploration of real content.

We implemented \sysName as a web-based system using a GPT-powered agent and integrated with Microsoft Power BI. 
The system supports any Power BI dashboard with appropriate metadata annotations.
It uses OpenAI's Realtime API~\cite{realtime} for speech processing and provides real-time visual feedback through dashboard highlighting and overlays.
Using the prototype, we conducted a mixed-methods user study comparing \sysName against a baseline system and collected both quantitative and qualitative data from participants with novice- to expert-level dashboard experience.
The study design incorporated insights from four pilot studies to refine the experimental protocol.
Results demonstrate that participants with \sysName had higher task accuracy as compared to the baseline.
They rapidly grasped \sysName's capabilities and consistently preferred multimodal interactions over single-modality approaches for onboarding tasks.
Visual highlighting combined with voice-based questions and responses emerged as the dominant interaction pattern.
All experts expressed strong interest in using \sysName for their professional dashboard work.

The contributions of this work are threefold:
(1) the concept of a multimodal dashboard onboarding assistant that combines natural language, visual highlighting, and gestural interaction to replicate human-expert guidance;
(2) the implementation of \sysName, demonstrating how large language models can be integrated with dashboard environments to provide contextual, situated assistance; and
(3) an exploratory study comparing \sysName against a standard onboarding procedure (static dashboard user guide), focusing on differences in task correctness, interaction strategies, and perceived onboarding experience.

\section{Related Work}
\label{sec:rw}

Here we review prior work on visualization dashboards, onboarding, multimodal interaction for data visualizations, and automatic approaches for dashboard generation.

\subsection{Visualization Dashboards}

\textit{Visualization dashboards} have become ubiquitous tools for data exploration and decision-making across diverse domains, yet their complexity often creates significant barriers for users.
Few's book on visualization dashboards~\cite{few_information_2006} outlined the many fine points and varied purposes of dashboard design in industry.
Sarikaya et al.~\cite{a_what_2018} provided a comprehensive survey of what the scientific visualization community means by ``dashboards,'' revealing their diverse applications and the varied expectations users bring to these systems.
Their work highlighted the fundamental challenge: dashboards must balance powerful functionality with usability.

Multiple studies have documented the struggles users face when working with dashboard systems.
Tory et al.~\cite{tory_finding_2022} examined how users attempt to find their ``data voice'' when working with dashboards, documenting persistent difficulties in navigation, interpretation, and effective use of interactive features.
Their findings reveal that even well-designed dashboards can overwhelm users who lack sufficient training.
Similarly, Walchshofer et al. \cite{walchshofer_transitioning_2023} observed the socio-technical challenges that arise when organizations transition to new dashboarding systems, noting that successful adoption depends heavily on both technical design and organizational support.

The visualization community has responded to these challenges by developing design frameworks and best practices.
Bach et al.~\cite{bach23dashboardpatterns_tvcg} identified recurring dashboard design patterns, providing a vocabulary for discussing effective dashboard layouts and interaction paradigms.
Lin et al.~\cite{lin24} developed DMiner, which provides design recommendations for dashboards based on analysis of a large dataset of existing dashboards.
Their system helps authors create more effective dashboards by suggesting layouts, chart types, and design patterns that have proven successful in similar contexts, representing a shift toward data-driven dashboard design that could reduce user confusion from the outset.

Despite these advances in design guidance, the challenge remains: complex dashboards with multiple coordinated views and rich interactions require significant investment of time and effort by users to master.
This persistent gap between dashboard capabilities and user proficiency motivates the need for more sophisticated onboarding and assistance mechanisms that can help users navigate these powerful but complex systems.

\subsection{Onboarding for Dashboards}

Given the popularity of visualization dashboards, \textit{visualization onboarding} has recently emerged as a research area, though solutions remain fragmented across different contexts and applications.
Stoiber et al.~\cite{c_visualization_2019} first characterized the onboarding space for individual visualizations, establishing foundational concepts, such as \textit{when}, \textit{where}, and \textit{why} to provide onboarding.
\sysName uses these concepts to provide an onboarding (\textit{where}), before and during a task (\textit{when}), and only when the user asks for it (\textit{why}).
They also identified the need for systematic approaches to help users learn visualization.
Their work highlighted various onboarding methods ranging from simple tooltips to more sophisticated guidance.

Building on these foundational concepts, Dhanoa et al.~\cite{v_process_2022} proposed a systematic process model for dashboard onboarding that identifies key stages, such as interaction with the onboarding system, manipulation of the dashboard content, and its presentaion with the help of artifacts to the users.
This framework helps understand onboarding as a structured process rather than ad-hoc training.
\sysName uses this process to specify the interactions, content and the artifacts that are presented to the user (shown in Figure~\ref{fig:teaser}).

Recent advances have focused on automating onboarding creation using tours.
Effective guided tours balance navigational constraints and freedom to facilitate learning~\cite{DBLP:conf/chi/ElmqvistTT08}.
Dhanoa et al.~\cite{DBLP:journals/tvcg/DhanoaHFEGS25} introduced D-Tour, a system for semi-automatically generating interactive guided tours for visualization dashboards.
Their approach reduces the manual effort required from dashboard authors while maintaining the ability to create engaging onboarding experiences.
Similarly, Hoque and Sultanum~\cite{hoque25dashguide} developed DashGuide, which enables the creation of interactive dashboard tours with minimal author input while allowing for manual refinement when needed.

The comparative evaluation of different onboarding approaches has revealed important insights about their effectiveness.
Stoiber et al.~\cite{stoiber_comparative_2022} systematically compared various onboarding methods, finding that interactive approaches generally outperform static documentation, though the optimal approach depends on user characteristics and dashboard complexity.

Despite these advances, most existing onboarding solutions remain primarily visual and static, relying on annotations~\cite{m_annotating_2012}, tooltips, or guided tours that users follow passively.
Chundury et al.~\cite{DBLP:journals/ivs/ChunduryYCMSE23} represent a notable exception with their contextual, in-situ help features for visual data interfaces, though their approach focuses on help rather than comprehensive onboarding.
This gap between the interactive nature of modern dashboards and the static nature of most onboarding solutions motivates the need for more dynamic, responsive onboarding systems that can adapt to user behavior and provide assistance through multiple interaction modalities.
Current automated onboarding methods are still far inferior to the gold standard of onboarding: using a human instructor.

\subsection{Multimodal Interaction with Data Visualizations}

The visualization community has long recognized the limitations of traditional mouse and keyboard interaction for complex data exploration tasks.
Elmqvist et al.~\cite{DBLP:journals/ivs/ElmqvistMJCRJ11} articulated the concept of \textit{fluid interaction} for information visualization, arguing that effective visualization systems should support seamless transitions between different interaction modes and reduce cognitive barriers between user intent and system response.
This foundational work established fluidity as a key principle and identified the utility of multiple modalities for designing more natural visualization interfaces.
Multimodality also has the benefit of supporting accessible visualization for non-visual audiences~\cite{DBLP:journals/tvcg/ChunduryPRTLE22}.

Lee et al.~\cite{lee12beyond} provided a comprehensive call to action, encouraging the visualization community to expand beyond mouse and keyboard to support more natural interactions.
Their survey of alternative interaction modalities highlighted the potential for touch, speech, gesture, and other modalities to make visualization more accessible and expressive.
This work catalyzed subsequent research into multimodal visualization interfaces.

Early explorations of touch-based visualization interaction demonstrated the potential of direct manipulation using modalities beyond the mouse and keyboard.
Baur et al.~\cite{DBLP:conf/tabletop/BaurLC12} introduced TouchWave, showing how kinetic multi-touch gestures could enable fluid manipulation of hierarchical stacked graphs.
Their work revealed how touch affordances could support more intuitive interactions with complex visualization structures.
Building on these foundations, Nielsen et al.~\cite{DBLP:conf/ozchi/NielsenEG16} studied the use of touch and pen scribbling to support multidimensional visualization queries.
Badam et al.~\cite{Badam2017} analyzed the affordances of different input modalities---touch, speech, proxemics, gestures, gaze, and wearables---for visual data exploration, providing guidelines for combining modalities to leverage their complementary strengths.
Thompson et al.~\cite{DBLP:conf/avi/ThompsonSS18} developed Tangraphe, which employed single-hand, multi-touch gestures for network visualization exploration, demonstrating how naturalistic gestures could replace conventional pointer-based interactions.

The integration of speech with visualization interaction has proven particularly promising for complex analytical tasks.
Srinivasan and Stasko~\cite{DBLP:conf/vissym/SrinivasanS17} examined what users naturally want to express through natural language when analyzing data, providing insights into the linguistic patterns that visualization systems should support.
They later developed Orko~\cite{DBLP:journals/tvcg/SrinivasanS18}, one of the first systems to combine voice and direct manipulation for network analysis, showing how speech could complement visual interaction for complex exploration tasks.
%
%More sophisticated natural language interfaces emerged with systems like Eviza~\cite{DBLP:conf/uist/SetlurBTGC16}, which provided comprehensive natural language support for visual analysis workflows.
These systems demonstrated that users could express analytical intents through conversational interfaces, though challenges remained in handling ambiguous queries and maintaining context across extended analysis sessions.
Building on these foundations, Srinivasan and Setlur~\cite{srinivasan2023bolt} developed BOLT, a natural language interface specifically designed for dashboard authoring, showing how conversational interaction could extend beyond data exploration to visualization creation itself.

Empirical studies have provided crucial insights into user preferences and performance with different modality combinations.
Saktheeswaran et al.~\cite{saktheeswaran20} found that participants preferred combining multimodal over unimodal input when exploring network visualizations on large vertical displays, with touch and speech proving to be a particularly effective combination.
Their findings challenged assumptions that users would prefer single-modality interaction for simplicity.
In large display contexts, Le{\'o}n et al.~\cite{molina24} found that participants preferred speech commands for 10 of 15 exploration tasks over touch, pen, and mid-air gestures, suggesting that speech excels particularly in scenarios where physical interaction is challenging.

Recent work has focused on designing consistent multimodal experiences across different devices.
Jakobsen et al.~\cite{DBLP:journals/tvcg/JakobsenHKH13} explored how user position and movement relative to large displays could inform interaction design, showing how proxemic cues could enhance information visualization experiences through implicit input; 
Badam et al.~\cite{DBLP:conf/ieeevast/BadamAEI16} extended this to multiple users.
Srinivasan et al.~\cite{srinivasan20inchorus} developed InChorus, which provides consistent multimodal interactions across tablet devices, demonstrating how design principles can maintain usability while adapting to different form factors.

\subsection{Automatic Generation of Dashboards}

The complexity of creating effective dashboards has motivated significant research into automated generation approaches.
Early systems such as VizDeck~\cite{DBLP:conf/sigmod/KeyHPA12} introduced the concept of self-organizing dashboards that automatically recommend visualizations based on statistical properties of data, using a card game metaphor to help users organize visualizations into interactive dashboards with minimal programming effort.

Deep learning approaches have shown promise for dashboard recommendation and generation.
Wu et al.~\cite{DBLP:journals/tvcg/WuWZHZQZ22} developed MultiVision, which uses deep learning models to assist in analytical dashboard design by learning from provenance data and authoring logs. 
Their mixed-initiative system allows users to provide optional input while the model recommends data column selections and multiple chart combinations.
Similarly, Deng et al.~\cite{DBLP:journals/tvcg/DengWQW23} proposed DashBot, which uses deep reinforcement learning to generate insight-driven dashboards by constructing training environments based on visualization knowledge.

Intent-based recommendation systems have emerged to support dashboard composition workflows. 
Elshehaly et al.~\cite{DBLP:journals/tvcg/ElshehalyRBMAGR21} developed QualDash for healthcare quality improvement, introducing a metric card metaphor that serves as a building block for generating adaptable dashboards across different hospital units.
Pandey et al.~ \cite{DBLP:journals/corr/abs-2208-03175} introduced \textsc{MEDLEY}, which recommends dashboard collections based on analytical intents such as measure analysis, change analysis, and distribution analysis.
Users can specify intents explicitly or implicitly through data attribute selection, with the system providing collections of logically grouped views and filtering widgets.

The proliferation of large language models has enabled new approaches to dashboard creation through natural language interaction.
Dibia~\cite{DBLP:journals/corr/abs-2303-02927} developed LIDA, which uses LLMs for automatic generation of grammar-agnostic visualizations and infographics, demonstrating how conversational interfaces can simplify visualization creation.
Building on this foundation, Lisnic et al.~\cite{lisnic25plume} developed a LLM-powered system specifically for creating text content within dashboards, showing how artificial intelligence (AI) can assist with narrative elements that provide context and explanation.

Recent work has integrated LLMs into broader visual analytics workflows.
Chen et al.~\cite{chen25interchat} developed InterChat, which combines direct manipulation with natural language interaction for visual analytics, including LLM-based prompting capabilities.
Zhao et al.~\cite{DBLP:journals/tvcg/ZhaoWXZGTZC25} proposed LightVA, a framework that uses LLM agents for task planning and execution in visual analytics, employing a recursive process with planner, executor, and controller components.
Wen et al.~\cite{wen25prompt} explored multimodal prompting with LLMs for visualization authoring, investigating how different input modalities can enhance the creation process beyond text-based interaction.

While these automated generation approaches have made dashboard creation more accessible, most focus on the authoring process rather than helping users understand and navigate existing dashboards.
The challenge of onboarding users to complex, pre-existing dashboard systems---particularly through multimodal interaction---remains largely unaddressed by current automatic methods.

\section{\sysName: Dashboard Interactive Assistant for Navigation and Analysis}

\begin{figure}[tbh]
  \centering
  \includegraphics[width=\linewidth]{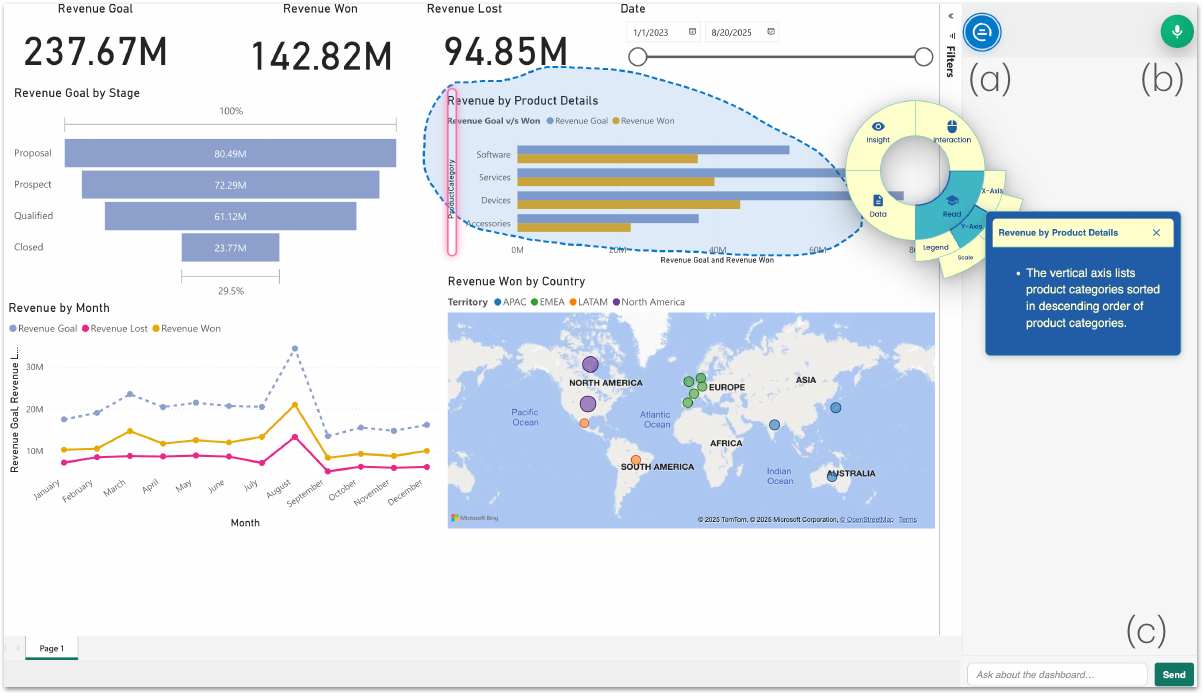}\\
  \includegraphics[width=\linewidth]{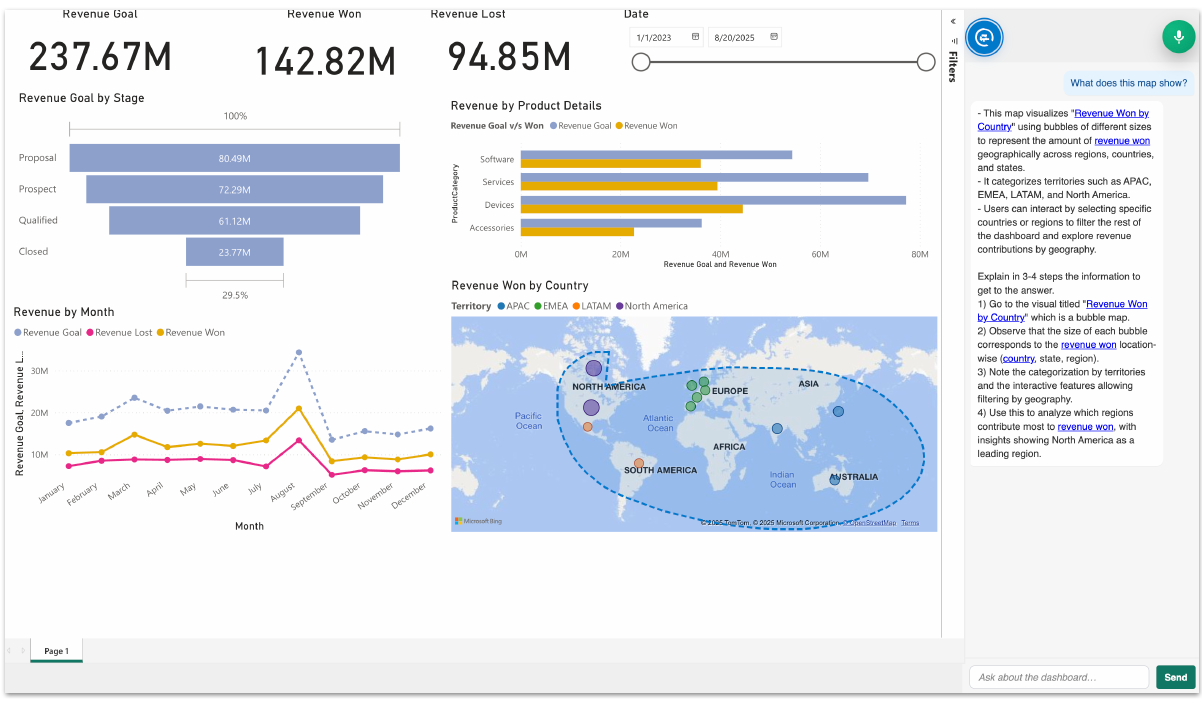}
  \caption{\textbf{\textsc{Diana} user interface.}
  \textnormal{The first image shows a visualization dashboard with Diana activated: (a) represents the lasso selection button, (b) push-to-talk for speech-based interaction, and (c) chat-based interaction.
  Currently, the lasso selection is activated and the y-axis of the visualization is highlighted, along with the radial menu which is opened for the y-axis information.
  The second image shows multimodal interaction with mouse-based lasso selection and chat-based interaction that infers the context of the conversation based on the selected visual.}}
  \label{fig:ui-side-by-side}
  \Description{The interface displays key revenue metrics along the top, followed by a funnel chart, a monthly line chart, a horizontal bar chart of revenue by product category, and a world map showing geographic revenue distribution. A tooltip explains the sorting of the vertical axis in the bar chart. The pattern shown across the dashboard is that the revenue goal exceeds the revenue won by a significant margin, the funnel chart demonstrates a typical narrowing from Proposal to Closed stages, and the monthly line chart suggests seasonal variability with a noticeable rise late in the year. The bar chart indicates that Software and Services are the strongest revenue categories, while Accessories generate the least. The world map highlights North America and Europe as the most revenue-dense regions, reinforcing the concentration of performance in specific geographic areas. The second figure has the emphasis on guided interpretation: the help panel clarifies how to read the map and highlights the contrast between regions, demonstrating how system-generated explanations scaffold the user’s understanding of broader revenue patterns.}
\end{figure}
Existing onboarding methods for dashboards, such as documentation, training, videos, and meetings, often fail to support users in situ.
They are static, linear, and cannot adapt to individual preferences or questions that may arise during the onboarding process. Additionally, the user needs to constantly switch programs, and manually test interactions while cross-referencing instructions, which can lead to frustration with onboarding help.
Self-designed tutorials, such as step-by-step interfaces, also face the challenge of being ignored by many users.
To address the balance between autonomy and agency, we designed \sysName (Dashboard Interactive Assistant for Navigation and Analysis), a multimodal onboarding assistant powered by Large Language Models.

\sysName is designed not to replace domain expertise, but to bridge the gap between static onboarding and in-situ analysis.
Its role is to help users navigate dashboards, understand visualizations, and learn how to perform analysis through guided, multimodal interaction. 
Users can interact with \sysName via three interaction modalities: keyboard (chat), mouse (lasso selection), and voice (speech).
Based on the chosen modality, \sysName provides feedback through text, audio, a contextual menu, and visual highlights to orient users.

Our design goals (DG) combine lessons from the state of the art in digital assistants (e.g., Microsoft Co-Pilot), common modalities already available in dashboards, and practices of human onboarders:

\begin{description}
    \item[DG1] Build on \textbf{familiar environments}: create an onboarding system that can be used on a familiar dashboard platform (such as Microsoft Power BI) to increase the likelihood of adoption.
    \item[DG2] Support \textbf{multimodal interaction}: allow users to interact via keyboard, mouse, and voice, individually or in combination, to accommodate different preferences and contexts.
    \item[DG3] Provide \textbf{human-like onboarding} support: simulate the role of an expert by answering questions, guiding navigation, and scaffolding analytical steps.
    \item[DG4] \textbf{Facilitate orientation} in complex dashboards: employ features such as lasso selection, contextual help menu, and visual highlights to direct user attention and clarify relationships across views.
    \item[DG5] Balance \textbf{guidance and autonomy}: give users the freedom to request help when needed, rather than forcing a step-by-step walkthrough.
\end{description}

We designed \sysName according to these goals that support the user's orientation and onboarding within the dashboard.
It is implemented using a Microsoft Power BI dashboard embedded in a custom web application, thereby satisfying \DGOne.
We distinguish between modalities, which define how users interact with \sysName (keyboard, mouse, voice), and features, which define how the system provides feedback and orientation (contextual help menu, visual highlighting).
We explain them below.

\subsection{Mouse-based Interaction}

The mouse is the primary input device in most dashboard environments, and users are generally familiar with interacting with visualizations through clicking.
To extend this familiar mode of interaction for onboarding (\DGTwo), we included two complementary features: lasso selection (or free-form selection) and hover functionality \inserted{(see Figure \mbox{\ref{fig:ui-side-by-side}} (a))}.
Lasso selection enables users to mark a region of interest within the dashboard, mimicking the pointing gestures a human expert might use when directing attention to a particular element.
Once a region is selected, the system captures the rectangular coordinates of the highlighted area and maps them to the corresponding visualizations.
At the same time, the dashboard remains fully interactive for further exploration. 

Hover functionality provides additional detail in two ways: (i) through contextual help options in the radial menu (see Section~\ref{sec:contextmenu}), and (ii) through embedded hyperlinks returned in responses to text-based queries (see Section~\ref{sec:keyboard}).
Therefore, lasso serves as a way of anchoring questions to a region, while hover enriches that region with further information.
In a typical onboarding scenario, users also point and ask questions. Mouse-based interactions allow us to provide human-like onboarding support (\DGThree) and also facilitate orientation by anchoring onboarding interactions to specific areas of the visualization (\DGFour).

\subsection{Voice-based Interaction}

With the growing support for multimodal interaction in LLM-powered assistants (e.g., Microsoft Copilot), we included a voice modality to mimic the experience of conversing with a human onboarder (\DGTwo and \DGThree).
Speech input allows users to issue queries in natural language and receive real-time guidance without relying on typing.

Users can enable this modality via a push-to-talk model\inserted{(see Figure \mbox{\ref{fig:ui-side-by-side}} (b))}, where they press and hold the mouse button while speaking, and upon release, \sysName processes the input using the Realtime API~\cite{realtime} and immediately returns a spoken response (\DGFive).
This design ensures that users maintain full control over when the system listens, avoiding interruptions during regular conversation.
We initially experimented with a wake-word approach (\textit{``Hey dashboard!''}), but found it prone to false activations, as the system occasionally responded to background discussions that were not intended for the assistant.
We therefore selected the push-to-talk model as a more reliable and less intrusive interaction style. 
In addition to the audio output, \sysName also highlights specific regions of the dashboard to direct the user's attention (\DGFour).

\subsection{Keyboard-based Interaction}
\label{sec:keyboard}

The keyboard-based interaction\inserted{(see Figure \mbox{\ref{fig:ui-side-by-side}} (c))} complements the voice-based interaction by offering a written chat interface that supports deliberate, text-based interaction.
Unlike spoken dialogue, written exchanges create a persistent trace of the conversation, allowing users to revisit instructions and follow them step by step (\DGTwo and \DGThree).

The chat component uses the OpenAI APIs to process user queries and returns responses in text. 
In addition to plain text, these responses may also include embedded hyperlinks, and hovering over these responses highlights the corresponding regions on the dashboard (e.g., the visualization, its legends, or its axes) (\DGFour and \DGFive).

\subsection{Contextual Help Menu}
\label{sec:contextmenu}

To support onboarding in a deterministic and structured way, we designed a contextual radial menu that complements the natural language modalities (chat and speech), which rely heavily on LLMs for generating responses.
While conversational interaction adapts to the user’s phrasing and provides guidance step by step, the menu makes the full range of onboarding information visible, even if the user does not know what to ask \cite{ko2004designing}.

We extended the component graph for dashboards described by Dhanoa et al.~\cite{v_process_2022} for the design of our contextual help menu.
As a visualization can contain many elements, such as data, legends, axes, and interaction options, that may require explanation, we adopted a bottom-up approach to assemble this information hierarchically.
To present this information visually, we used a circular radial layout (CRL) that progressively reveals detail across levels. 
Pourmemar et al.~\cite{menus} found that using the CRL was faster than the hierarchical drop-down menu in an immersive augmented reality environment.
Samp and Decker also found that pointing is faster with CRL and it is beneficial for the design of displayed hierarchical menus~\cite{10.1145/1842993.1843021}.

Therefore, we also used CRL to capture the complexity of the component graph in a structured and navigable form.

The menu is organized into four top-level categories:

\begin{itemize}
    \item \textbf{Read}, describing how to interpret the visual encoding (e.g., axes, labels, scales, legend order).

    \item \textbf{Data}, explaining underlying sources such as tables, columns, keys, and measures.

    \item \textbf{Interact}, making explicit both self-interactions (e.g., highlighting or comparing categories) and cross-visual interactions (e.g., filtering other charts).

    \item \textbf{Insight}, outlining descriptive and smart insights such as trends or potential drivers.

\end{itemize}

The menu appears automatically when a region of a chart is selected with the lasso (\DGFour).
It opens at the relevant category and displays a pop-up information box with a description of the selected region.

When users hover over menu items, the information box displays the corresponding explanation.

Although the schema is deterministic, we used an LLM to polish the narrative text associated with each menu entry.
For instance, instead of a bare label such as \textit{Currency Scale}, the menu might display \textit{The X-axis represents values in USD on a continuous scale.}
In this way, the LLM enriches the text with more natural explanations, while the structure and content remain grounded in the underlying visualization metadata.
In addition to plain text, we use icons to enhance readability.

The menu dynamically adapts to the visualization type by pruning irrelevant categories.
For instance, for a Key Performance Indicator (KPI) card, interaction options are hidden, as direct highlighting or filtering is not possible.
Similarly, for a filter visual, the Insight category is suppressed, since filters do not directly encode data-driven findings.

Unlike the chat or speech modalities, which reveal information only when prompted, the menu lays out the full structure of what can be learned, encouraging discoverability and providing a complete description of the visualization.
It packages information by category but requires the user to assemble the pieces into an understanding of the visualization.

\subsection{Visual Highlighting}

We designed the visual highlighting feature to complement all three modalities by directing user attention to specific dashboard elements, similar to how a human onboarder might point to a chart or annotate it (\DGThree).

When a user lasso-selects a region, the system maps its rectangular coordinates to identify which visualization components they intersect.
For keyboard or voice-based interactions, it extracts keywords (e.g., title, axis, legend) and maps them to the corresponding visualization components based on words in the user’s input (written or spoken).

These highlights appear as pulsating rectangular bars that serve as visual cues, helping users orient themselves and understand where to focus (\DGFour).

During a lasso selection, for example, if the system detects a region such as a legend, the highlight is shown around the legend area.
For voice-based interaction, highlights are synchronized with the spoken guidance, though slight rendering delays can occur. 
In keyboard-based interaction, highlights are triggered through embedded hyperlinks and appear when hovered.

\subsection{Combining Modalities and Features}

In real onboarding scenarios, users rarely rely on a single modality when seeking onboarding support. 
Therefore, we designed \sysName to allow users to combine modalities and features, in parallel or sequentially, to reproduce an onboarding experience similar to that provided by a human onboarder (\DGThree).
Our system enables users to lasso a region (mouse-based interaction) and simply use the voice-based interaction to ask a question for the selected area.
The user can also view more detailed information in the contextual menu, while receiving audio onboarding.
The use of visual highlights can prompt further exploration, leading the user to open the contextual menu for that visualization or simply select it and ask a chat-based query (keyboard interaction).

These combinations allow \sysName to provide different kinds of support: mouse-based anchoring to a selected region, a contextual menu for structured information, natural language support via voice and keyboard-based interactions, and visual highlighting for orientation and directed attention.
Altogether, they can create an adaptive experience in which onboarding can be tailored to both the task at hand and the user’s preferred way of interacting.

Unlike typical LLM-based assistants that aim to deliver direct answers, \sysName focuses on orientation and learning.
The system is prompted with information about the dashboard's structure (such as available visuals, axes, maps, and overview descriptions) but not with the underlying data values.
As a result, \sysName cannot return analytical results itself, except for basic descriptive queries (e.g., \textit{what is the y-axis?}).
This design choice ensures that the assistant focuses on orientation and guidance rather than replacing the user’s own analytical work.
We next explain how \sysName works.

\section{Implementation}

We implemented \sysName using the Microsoft Power BI Embedded report~\cite{microsoft_embedded} to integrate our solution within an existing commercial dashboard tool such as Microsoft Power BI. 
We obtain information about all visualizations in the dashboard using Power BI's REST API~\cite{microsoft_restapi}. We also apply simple heuristics to deduce information not exposed through these APIs, such as determining the positions of legends and axes, which are not provided by default.
We infer these by adding padding offsets to the top-left rectangular coordinates of each visualization based on its type.
This allows us to construct an internal map of the dashboard layout and the relative placement of its components.
As different visualization types place their axes in different locations, we extract those positions as well by checking them against specific Power BI keywords.
We also create a transparent overlay on the dashboard for lasso selections.
When a user draws a region, we capture its bounding box and map it onto the dashboard layout.
The system itself is built using the Vue framework~\cite{vue} for the frontend and Flask for the Python backend.

Using Vue, we implement dedicated components for all modalities and features, including visual highlights and the radial menu.
When a user selects a region, the system captures its coordinates and maps them onto the dashboard layout to answer pointed questions. It then activates the appropriate Vue components, such as those for visual highlighting and the radial menu, which are rendered as overlay layers on top of the embedded Power BI iFrame.

When the user interacts through another modality, such as the keyboard or speech, the system forwards the relevant visualization metadata to the backend. The backend then integrates this contextual information with our prompts to generate appropriate onboarding instructions. The system maintains this context until the user selects a different region or asks a new question.

For the natural language interaction, we use the OpenAI APIs: Chat Completions with gpt-4.1-mini for text~\cite{gptmini} and the Realtime API~\cite{realtime} with gpt-4o-realtime-preview for voice I/O (with automatic speech transcription).
All OpenAI calls are proxied through the Flask backend~\cite{flask}, which mints ephemeral Realtime session tokens so the browser never sees the primary API key.

\sysName cannot query the underlying dataset nor retrieve off-screen values.
Therefore, it cannot provide data-derived answers or numeric values.
It only uses the metadata and content available from the embedded dashboard (e.g., visual titles, layout bounds, and author-provided descriptions).
\sysName has no authoring control over the dashboard and therefore cannot apply filters or interact with visuals.
When a user request requires information that is not available from the visible dashboard context, the system emits a structured “NOT IN DATA” outcome and redirects the user to the appropriate navigation or exploration steps.

\section{Study Design and Methodology}

Our goal was to examine how users engage with \textsc{\sysName} (\DGTwo and \DGThree) and how it affects the balance between guidance and user autonomy (\DGFive).
To provide an appropriate baseline for comparison, we also included a condition where participants were not given access to \sysName at all and could only use a standard static dashboard guide provided as a printed document.

We therefore adopted a mixed-methods approach, measuring task performance across both conditions and conducting qualitative analysis of think-aloud sessions and subsequent semi-structured interviews.
We conducted the study in two phases:
(1) a formative study (n=4), to refine the task design and the set of supported modalities, and
(2) main interviews with 12 participants, including six participants using \sysName (\DianaParticipant{1}-\DianaParticipant{6}) and six participants using a static dashboard guide (\BaselineParticipant{1}-\BaselineParticipant{6}).
The participants ranged from novice to experienced dashboard users.

During the main sessions, participants completed a series of dashboard-oriented tasks of varying difficulty, either using \sysName or the static dashboard guide.

All (n=12) participants received the static dashboard guide, but only half (\DianaParticipant{1}-\DianaParticipant{6}) had access to Diana in addition to the guide. 
Participants using \sysName were free to choose any interaction modality (or modalities) per task and could ask follow-up questions on tasks to \sysName if needed.

\subsection{Participants}

\begin{figure}
    \centering
    \includegraphics[width=\linewidth]{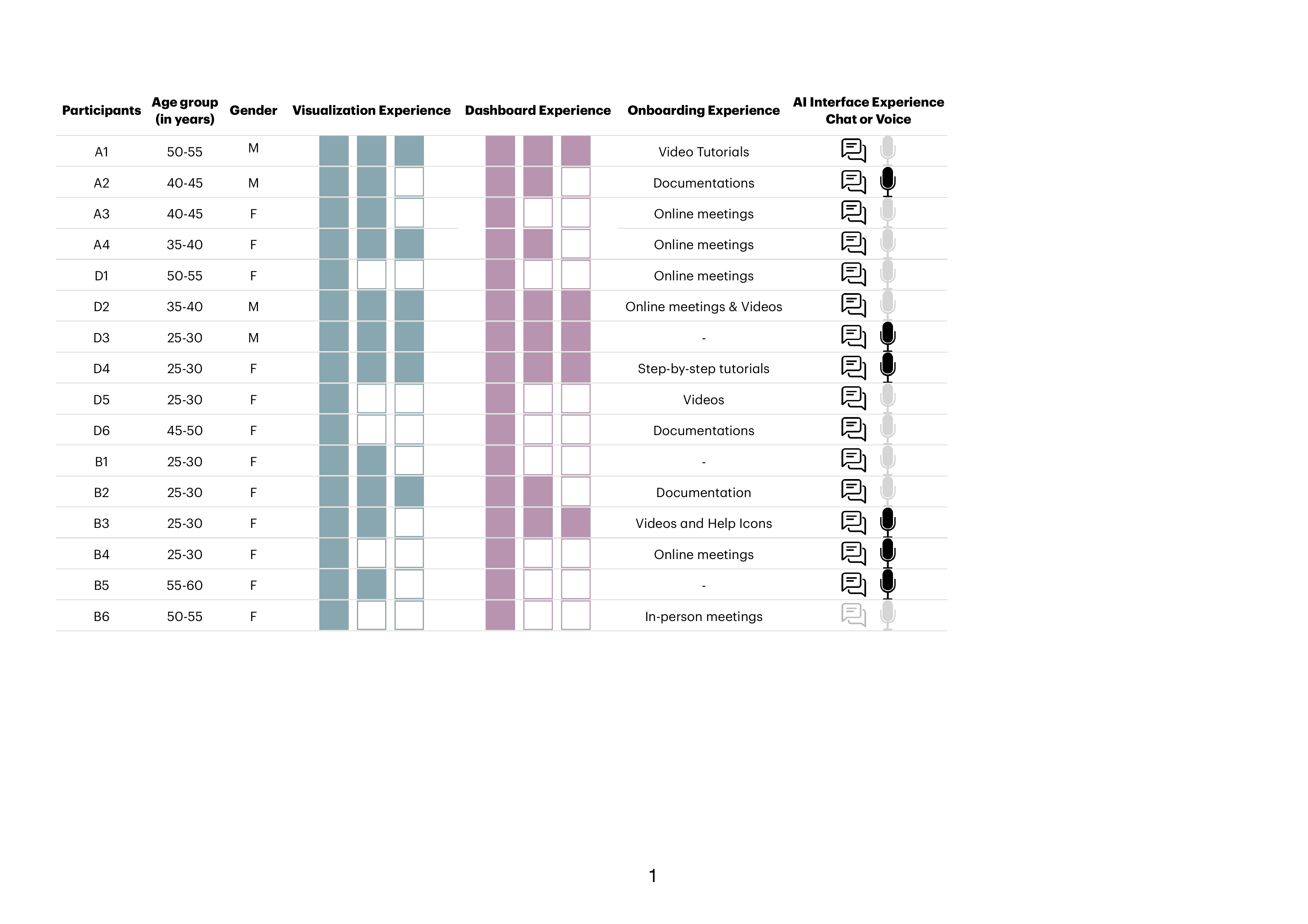}
    \caption{\textbf{Participant demographics.}
    \textnormal{Overview of the formative study participants (A1-A4), \sysName condition participants (\DianaParticipant{1}-\DianaParticipant{6}, and baseline condition participants (\BaselineParticipant{1}-\BaselineParticipant{6}).
    Gender is encoded as male (M) and female (F).
    Visualization and dashboard experience is measured from novice to expert.
    AI interface experiences are noted for chat and voice interfaces.}}
    \label{fig:demographics_table}
    \Description{The demographics table lists each participant’s age, gender, visualization experience, dashboard experience, onboarding background, and prior AI interface experience. A pattern emerges in which younger participants, particularly those aged 25–30, tend to show higher visualization and dashboard experience as indicated by more filled bars, whereas older participants often exhibit less direct tool experience but may have engaged more with documentation or training materials. Chat-based AI experience is common across groups, but voice-based interaction appears less widespread. Differences also appear across participant groups: the A group frequently reports structured onboarding sources such as documentation or videos, the D group often has blended onboarding exposure, and the B group includes several participants with little or no formal onboarding experience.}
\end{figure}

We first conducted a formative study with four participants, out of which one was (A1) a dashboard author who regularly onboards several colleagues, and the other three (A2-A4) were domain experts and dashboard users with different levels of visualization and dashboard expertise, as shown in Figure~\ref{fig:demographics_table}.
We used the insights from the formative study to inform the task phrasing and overall improvement of the prototype for the main session.
We do not include the formative study data in the main analysis.

Both the formative study and main session participants were recruited from different university departments, such as Business Intelligence, Economics, Finance \& Controlling, and Computer Science.
For the main session, the average age of the participants was 38.4 years ($\sigma$ = 10.2), and their experience in their field of expertise ranged from 2 to 25 years. 
The first six participants (\DianaParticipant{1}-\DianaParticipant{6}) received \sysName, in addition to the standard user guide.
Among them, three were well-versed with dashboards (\DianaParticipant{2}, \DianaParticipant{3}, and \DianaParticipant{4}), with \DianaParticipant{2} providing regular onboarding to many of his colleagues.
The other three participants (\DianaParticipant{1}, \DianaParticipant{5} and \DianaParticipant{6}) were novices and had little to no experience with visualization dashboards.
Among the baseline participants (\BaselineParticipant{1}-\BaselineParticipant{6}) who received only the standard user guide, one had intermediate knowledge about the dashboards (\BaselineParticipant{2}), and one was an expert (\BaselineParticipant{3}).
The other four participants (\BaselineParticipant{1}, \BaselineParticipant{4}, \BaselineParticipant{5}, and \BaselineParticipant{6}) were novices but had some experience with visualizations.
Each interview was conducted in person and individually. 

We also asked all these participants about their familiarity with AI tools and their experience with different modalities.
These participants were chosen for two main reasons: (i)~to understand how varied level of expertise (novice to expert) leads to differences in self-onboarding practices, and
(ii)~how does experience with AI and multimodality inform user preferences during onboarding. 

\subsection{Dataset and Task}

We used an example dataset from Microsoft Power BI~\cite{examplebi} to create a custom visualization dashboard.
The dashboard (Figure \ref{fig:ui-side-by-side}) included three key performance indicators (KPIs), a date filter, a funnel chart, a line chart, a clustered bar chart, and a map visualization.
All visualizations, except the KPIs, were interactive.
In addition to highlighting and filtering, a few visuals supported drill-down, allowing users to explore information in greater depth. 

We leveraged this interactivity when designing the task set, which spanned task type, cardinality, and level of difficulty.  

\begin{itemize}
    \item \textbf{Task type:} Tasks were categorized into three groups: \emph{information lookup}, \emph{exploratory}, and \emph{interpretive}.
    Information lookup tasks ranged from simple retrievals (e.g., reading a KPI value) to slightly more involved actions (e.g., hovering over a visual to reveal details).
    Exploratory tasks required participants to interact with one or more visuals to uncover the relevant information.
    Interpretive tasks went further, requiring participants to make sense of patterns or relationships that were not explicitly presented.  

    \item \textbf{Cardinality:} This dimension captured the number of views a task required.
    As the dashboard consisted of multiple coordinated views, some tasks could only be solved by interacting with multiple visualizations.  

    \item \textbf{Level of difficulty:} We assigned difficulty levels (easy, medium, hard) based on our prior experience with dashboard onboarding and the complexity of the interactions required to solve the task, also informed by the formative study results.
\end{itemize}

For instance, an easy look-up task involving only a single view (Figure \ref{fig:ui-side-by-side}) could be \textit{How do I figure out the scaling of the x-axis of the bar chart?}
Meanwhile, a difficult task involving multiple views could be \textit{How can I figure out the revenue goal for Australia in the Services subcategory for the Proposal stage?}
We designed ten such tasks based on our own onboarding experience and in discussion with the formative study participant A1.
Figure~\ref{fig:study-results} shows the tasks, their types, and their difficulty levels.

\subsection{Study Procedure}

Each session began with an introduction to the study goals and a request for informed consent regarding audio and screen recording.
After receiving their consent, we asked participants to complete a pre-questionnaire containing preliminary questions about their prior experience with visualizations, dashboards, and AI interfaces (summarized in Figure~\ref{fig:demographics_table}). 

\paragraph{\sysName}

After the pre-questionnaire, the participants (\DianaParticipant{1}-\DianaParticipant{6}) completed a short training session lasting approximately five minutes.
During the training, we introduced \sysName on the dashboard and asked participants to try each interaction modality at least once and to experiment with combining modalities.
This was intended to ensure that participants felt confident in using \sysName before starting the tasks with the dashboard.
In addition to the modalities provided by the system, a dashboard user guide (as a digital document) was available throughout the study.

The task phase of this study lasted around 25 minutes.
Participants were free to choose whichever modalities they preferred for each task and could switch between them at any time.
They could ask the system as many questions as they wanted until they were satisfied, or solve the tasks on their own without assistance.

\paragraph{Static Dashboard Guide}

The static dashboard guide included text from the Microsoft Power BI sample dashboard webpage~\cite{examplebi}, which provides a brief overview of the visuals and how to interact with them.
In addition, drawing on our prior onboarding experience, we included a section that explains each visual, including its name, title, and an accompanying figure.
We also added the dashboard's purpose, a brief overview of available interactions, and general usage tips, resulting in a guide totaling 3.5 pages.
\BaselineParticipant{1} to \BaselineParticipant{6} received the standard user guide and were given time to review it and use it throughout the study.
They could choose whether to enter the task phase immediately or go through the documentation first. 
The task phase of this study lasted around 30 minutes.

In both study scenarios, if a task could not be solved, participants were allowed to skip it and continue to the next one.
After completing all tasks, participants were interviewed about their onboarding experience, comparisons with their usual onboarding practices, and preferred modalities (only for participants in the \sysName condition).

\subsection{Data Collection and Analysis}

To capture participants’ backgrounds and experiences, we collected data using the pre-questionnaire and follow-up interviews after the tasks, focusing on onboarding practices, modality preferences, and overall reflections on the experience.

For each session, we also recorded audio and screen activity, capturing both the dashboard interactions and the participants' think-aloud comments.
Responses to the pre-questionnaire and the follow-up interview were also documented.
The recordings were later analyzed qualitatively to identify task-solving strategies and modality choices.

For each task we recorded whether participants successfully completed it and whether they used the provided onboarding material or assistance.
For participants using \sysName, we also noted the modality, any switches between modalities, the features participants engaged with, and their rationales for switching.

After each task, we conducted brief interviews with the participants to gather qualitative feedback on their perceived level of guidance and autonomy.
For the \sysName group, we also asked about their preferred modalities and their experiences compared to prior onboarding approaches.
The time required to complete a task was not taken into account, as it depends heavily on participants' understanding of the tasks and their familiarity with the various modalities.

\section{Results}

In the following, we present the findings from the four formative study sessions and \sysName and static dashboard guide sessions.

\subsection{Formative Study Interviews}

We conducted a formative study with four participants (A1-A4) to refine both the design and implementation of \sysName and the experimental procedure. 
The list of participants included a Power BI specialist (A1), a regular Power BI user (A2), a dashboard author and occasional user (A3), and a daily Power BI user (A4).
In the following, we describe their experience with dashboards and onboarding, and their impressions of the system.

\textbf{A1} regularly created onboarding materials for dashboards, typically using videos and bookmarks in Power BI.
Their approach involved taking screenshots of dashboards, adding markers and pointers, and displaying them as a translucent overlay on the first page, which users can dismiss after viewing.
They noted that this process is time-consuming and requires frequent updates as dashboards evolve.
A1 also emphasized that dashboard features often contain subtle behaviors that are difficult to onboard effectively.
For example, drill-down interactions usually confused users because the view changes immediately, and users may not recognize the update.
During task performance, A1 suggested that visuals should be highlighted directly and that the AI should provide more specific guidance, while also conveying the dashboard's purpose.

\textbf{A2} relied heavily on dashboards in Microsoft Fabric and typically used official documentation for onboarding.
If documentation was insufficient, they turned to community forums, and only as a last resort, to AI systems.
Having used voice interfaces for more than seven years, A2 expressed disappointment that these technologies have not met their early promises, preferring instead what they described as ``ground truth'' onboarding methods.
They also noted that seeking help was particularly difficult for novices who may not yet know what to ask. A2 valued contextual menus that provided embedded information about a chart.
Based on this feedback, we added an additional onboarding option: a user guide with annotated images and instructions for reading, interacting with, and interpreting the dashboard.

\textbf{A3} used Power BI dashboards infrequently.
When joining her institution, she received a single 45-minute onboarding session covering relevant dashboards.
For support, she typically experimented on her own or asked a mentor.
During the tasks, A3 expected contextual menus to provide richer explanations.
Although she did not normally use voice interfaces, during the study, she quickly shifted to voice to ask clarification questions---about abbreviations, interaction methods, and task instructions.
She appreciated the multimodality of the onboarding, as she valued having options.
While she preferred chatting via text over talking out loud, she emphasized the importance of having a choice.
She also combined mouse-based selection with writing to find answers.

\textbf{A4} used dashboards daily and, like A3, had received only a single onboarding session.
When encountering issues, she usually consults her mentor.
Although she used Microsoft Co-Pilot for other tasks, she had not used it with dashboards.
During the tasks, she favored chat for its step-by-step guidance, but also experimented with speech interaction.
She appreciated that the Open AI's Realtime API recognized her accent and changed to a different language accordingly, but suggested improvements such as synchronized animations, slower speech, and a transcription option. 
The user guide, by contrast, felt burdensome.
A4 preferred using text and speech together: the chat allowed her to get written instructions while the voice output guided and confirmed her actions in real time.
She saw this combination as a way to save time for both herself and her mentor, reducing the need to ask experts for help.

Our formative study highlighted four changes:
(i) Visual highlighting is essential to direct attention and orient users within the dashboard,
(ii) prompts must be instructional so that they can teach a user how to perform concrete actions (e.g., drill-down on a visualization),
(iii) the information text box that appears with the contextual menu should have narrative style information with icons to enhance readability, and
(iv) a ground-truth fallback is needed, such as a  user guide so that the users can use it, if they prefer documentations over \sysName's support.

\subsection{\sysName Condition}
\begin{figure*}[t]
    \centering
    \includegraphics[width=\linewidth]{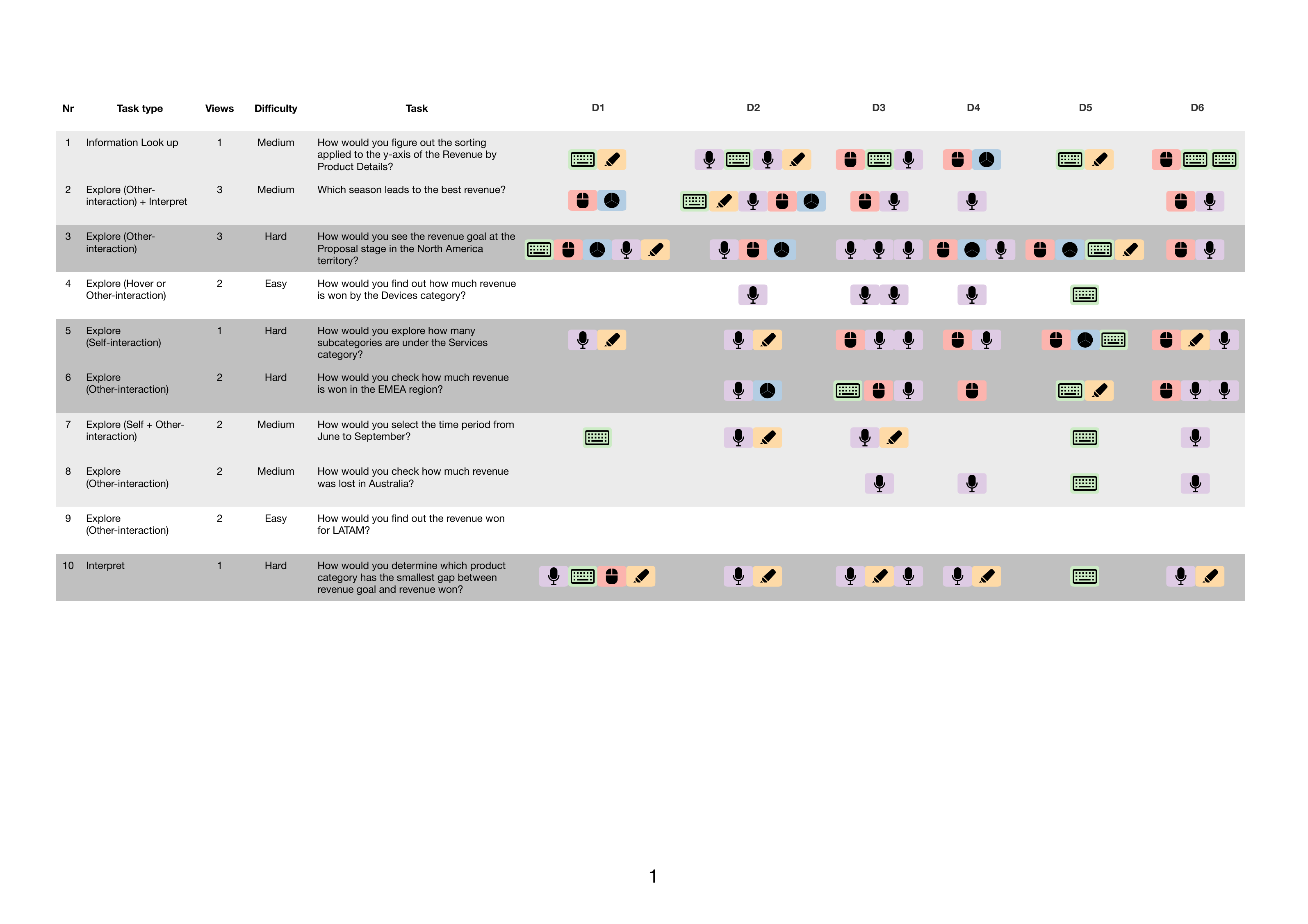}
    \caption{\textbf{Order of modalities and features used by \sysName participants during tasks.}
    \textnormal{Tasks are ordered by the order in which they were given to the participants.
    They are colored by their difficulty level: \colorbox{white}{easy}, \colorbox{medium}{medium}, and \colorbox{difficult}{difficult}.
    Modalities are represented through colors and icons: mouse \colorbox{mouse}{\faIcon{mouse}}, keyboard \colorbox{keyboard}{\faIcon{keyboard}}, and speech-based \colorbox{voice}{\faIcon{microphone}} interactions.
    Visual highlights \colorbox{highlight}{\faIcon{highlighter}} and radial menu \colorbox{radial}{\faIcon{chart-pie}} are also specified.
    The dashboard guide \faIcon{file} was not used by anyone and is not represented in the figure.
    An empty cell indicates that the participant solved the task on their own without the use of \sysName.}}
    \label{fig:study-results}
    \Description{This table presents ten tasks along with their difficulty, task type, and the interaction modalities used by each Diana participant. The overall pattern shows that Hard tasks consistently require more complex interaction strategies, with participants combining modalities such as hovering, clicking, typing, or voice commands to complete them. Easy tasks tend to rely on a single, simple interaction mode, usually involving only mouse actions. Medium tasks fall in between, with variability in whether participants resort to voice input or keyboard actions. Some participants repeatedly employ a wide range of modalities, while others rely on a narrower set, resulting in a visible correlation between task complexity and the breadth of interaction required.}
\end{figure*}

In both studies, participants completed the same set of dashboard-related tasks under two conditions: six participants used \sysName in addition to the standard user guide, and six used the standard user guide only.
Their performance varied by task type, number of views involved, and task difficulty (shown in Figure \ref{fig:study-results}).
The \sysName group frequently switched modalities as tasks grew more complex, whereas baseline participants relied solely on textual instructions and standard dashboard interactions.
We describe the study results below, first for \sysName and then for the static dashboard guide.
\subsubsection{Easy Tasks}

Most participants solved easy lookup or hover tasks independently, relying little on \sysName. 
Onboarding was mainly used for confirmation rather than discovery.
For instance, Task 4, which required looking up a \deleted{catgeory}\inserted{category} value in the bar chart, \DianaParticipant{1} solved it on her own as she is \textit{``used to figuring things out on [her] own,''} describing the multimodal help as \textit{``brand new... but pretty cool to have these options.''} (shown in Figure \ref{fig:study-results}).
Meanwhile, \DianaParticipant{2} and \DianaParticipant{3} used speech interaction and employed the voice and visual highlights feedback to double-check their answers, while \DianaParticipant{5} used the chat interface to clarify how to solve that task.

The learning effect was evident by Task 9 (shown as an empty row in Figure \ref{fig:study-results}), which required selecting a region on the map and reading the relevant value on hover or from one of the Key Performance Indicators (KPIs).
Every participant solved this task without any onboarding, regardless of prior expertise.

\subsubsection{Medium Tasks}

Medium-level tasks typically required interacting with two to three views and triggered frequent back-and-forth interactions with \sysName and switching heavily between modalities.
Participants \DianaParticipant{1} and \DianaParticipant{5} used the keyboard to chat with \sysName for most of the tasks.
The text-based response, coupled with embedded visual highlighting of the relevant region, helped participants understand where to look and how to execute the onboarding steps to complete the tasks (\DGFour).
While solving Task 1, which required a lookup and interpretation within the bar chart, \DianaParticipant{1} remarked that \textit{``it is pretty clever that it doesn’t tell you the answer, because sometimes you write [or type] and you get the answer, and you don’t know how you actually got there! That’s pretty cool.''}

For almost all medium tasks, \DianaParticipant{2}, \DianaParticipant{3}, \DianaParticipant{4}, and \DianaParticipant{6} used speech interaction, sometimes in combination with the mouse (\DGTwo).
For instance, in Task 2, which required locating and interpreting the answer in the line chart, \DianaParticipant{2} first used the chat interface to ask the question. 
They then used the visual highlights feature within the chat interface to orient themselves in the dashboard and then switched to the voice interface to ask the question again.
Since \sysName does not provide direct answers, they then used the radial menu to discover the insight needed to answer it.
The audio and visual feedback also helped \DianaParticipant{3}, \DianaParticipant{4}, and \DianaParticipant{6} orient themselves on the dashboard and find the right steps to get to their answers.
\DianaParticipant{2} also switched modalities, from keyboard to speech, to rephrase the question in a more natural-sounding way.
\DianaParticipant{1} and \DianaParticipant{2} also used the mouse to navigate the radial menu to learn more about the onboarding content.
This illustrates how participants used \sysName not only to retrieve answers, but also to scaffold their own orientation and reasoning in the dashboard.

\subsubsection{Hard Tasks}

Hard exploratory and interpretive tasks were the most demanding, often involving multiple steps and rephrasing.
Participants many times engaged in multi-modality sequences (e.g., mouse → voice → keyboard).
\DianaParticipant{2}, \DianaParticipant{3}, \DianaParticipant{4}, and \DianaParticipant{6} used the mouse and voice modality and rephrased the questions in their own way to interact with \sysName.
\DianaParticipant{2}, \DianaParticipant{3}, and \DianaParticipant{6} asked several follow-up questions using the voice modality, indicating that as tasks became more complex, the need for human-like onboarding became more prevalent (\DGTwo and \DGThree).
For instance, in Task 6, the question included the term EMEA, which is represented in the map visual on the dashboard.
\DianaParticipant{3} first used the chat interface to ask about the EMEA region.
However, this did not \inserted{yield a meaningful explanation and only provided onboarding steps for selecting} the region.
For this reason, he switched to the voice interface, which provided a clear definition of EMEA.

For the same task, \DianaParticipant{2} used the voice interface and then switched to the radial interface to verify the information obtained.
\DianaParticipant{5} and \DianaParticipant{6} switched between keyboard and other modalities.
Despite using all modalities, \DianaParticipant{1} could not solve Task 3, which required interacting with the map and the funnel visual on the dashboard.
She first rephrased the question in the chat interface, which returned a response indicating that the data was unavailable but provided onboarding instructions. 
However, the instructions did not explain how to select multiple points on the map.
She encountered similar issues when she switched from the radial menu to the voice interface.
This indicated that the instruction prompt to \sysName could have been improved.
This was the only unsolvable task during the entire user study session across all participants \inserted{in the \mbox{\sysName} condition}.

\subsection{Static Dashboard Guide (Baseline) Condition}

All the participants (\BaselineParticipant{1}-\BaselineParticipant{6}) went through the provided user guide before starting the tasks.
Based on their personal reading preference, a few participants (\BaselineParticipant{1}, \BaselineParticipant{3}, \BaselineParticipant{5}, and \BaselineParticipant{6}) simply glanced at the user guide, while others read the descriptions in detail (\BaselineParticipant{2} and \BaselineParticipant{4}).
\BaselineParticipant{4}  \deleted{went through the document thoroughly and even tried out various interactions listed in the guide in order}\inserted{thoroughly reviewed the document and even tried out various interactions listed in the guide} to familiarize herself with the interaction concepts.

\begin{figure}[htb]
    \centering
    \includegraphics[width=\linewidth]{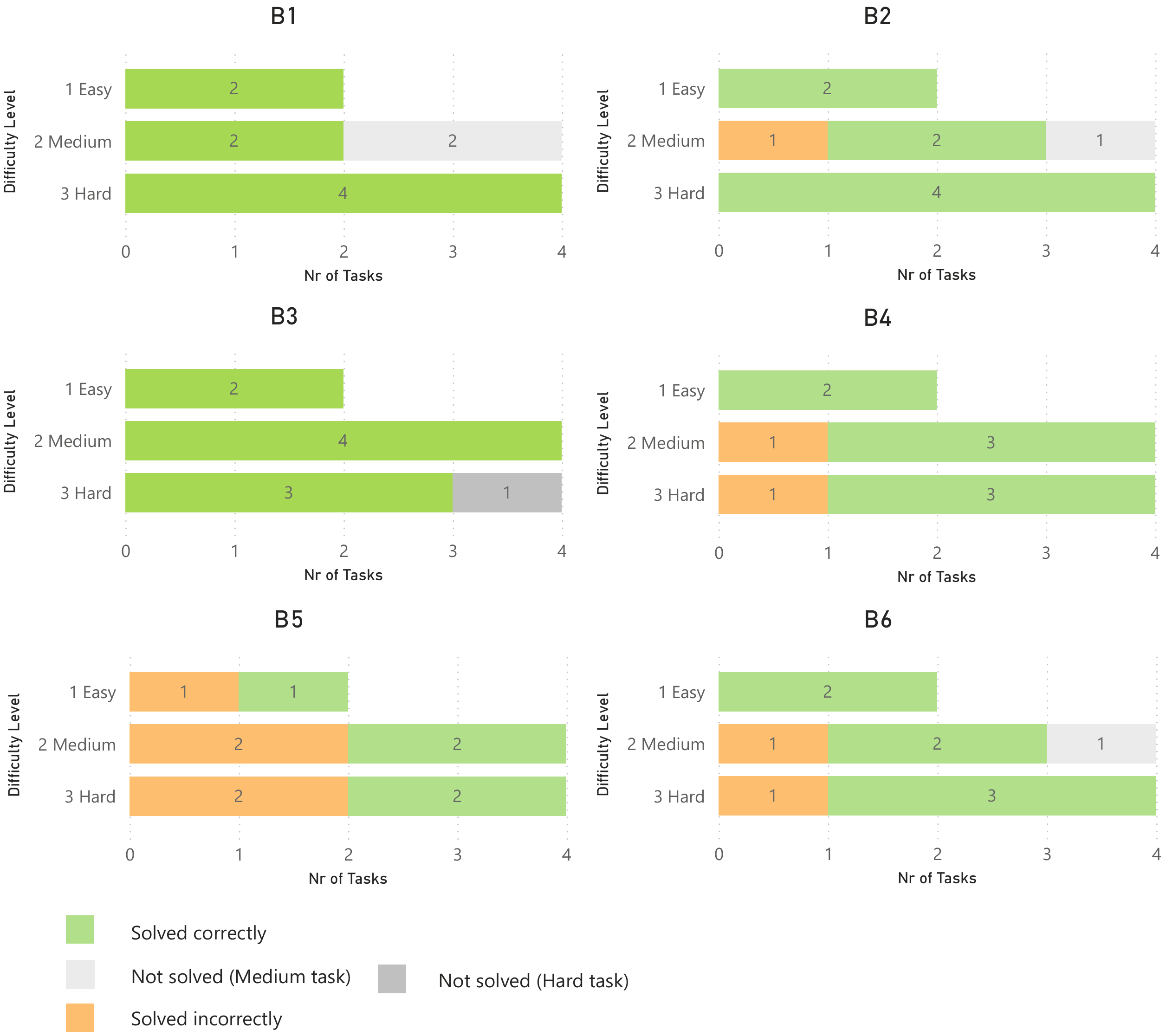}
    \caption{\textbf{Participant results from the baseline condition.}
    \textnormal{The results are shown for each participant (\BaselineParticipant{1} - \BaselineParticipant{6}) broken down into task difficulty and colored by task accuracy and completeness.}}
    \label{fig:baseline-results}
    \Description{The figure contains six bar charts labeled B1 through B6, each showing how many Easy, Medium, and Hard tasks were solved correctly, solved incorrectly, or not solved. Across all six charts, the overall pattern is that Easy tasks are most often solved correctly, Medium tasks show mixed outcomes with noticeable variation between participants, and Hard tasks consistently produce the greatest number of incorrect or unsolved attempts. The charts collectively illustrate a clear decline in accuracy as task difficulty increases, with each participant group exhibiting its own variation around this general trend.}
\end{figure}

\subsubsection{Easy Tasks}

Almost all the participants solved the easy tasks correctly.
However, four out of the six participants faced difficulties due to a combination of limited dashboard familiarity, the lack of comprehensive onboarding material, and not always reading or fully processing the provided instructions.
This led them to adopt workarounds.
For example, \BaselineParticipant{5} and \BaselineParticipant{6} performed manual calculations for Task 4 because they were unsure how to find the relevant summed values directly in the dashboard.
In the process, \BaselineParticipant{6} discovered the bar chart tooltip unintentionally and then used it to answer the question correctly.

For Task 9, participants \BaselineParticipant{3}, \BaselineParticipant{4}, \BaselineParticipant{5}, and \BaselineParticipant{6} manually calculated each data point, either mentally or using the provided pen and paper.
Despite these efforts, \BaselineParticipant{5} still gave an incorrect answer (as shown in Figure \ref{fig:baseline-results}) because she did not realize that the tooltip displayed both the actual and the highlighted values for each point.
This indicates that even basic dashboard interactions were not always transparent.

\subsubsection{Medium Tasks}

Medium-level tasks yielded more incorrect answers, and even participants who answered correctly often expressed uncertainty and sought verbal confirmation.

\begin{figure}[htb]
    \centering
    \includegraphics[width=\linewidth]{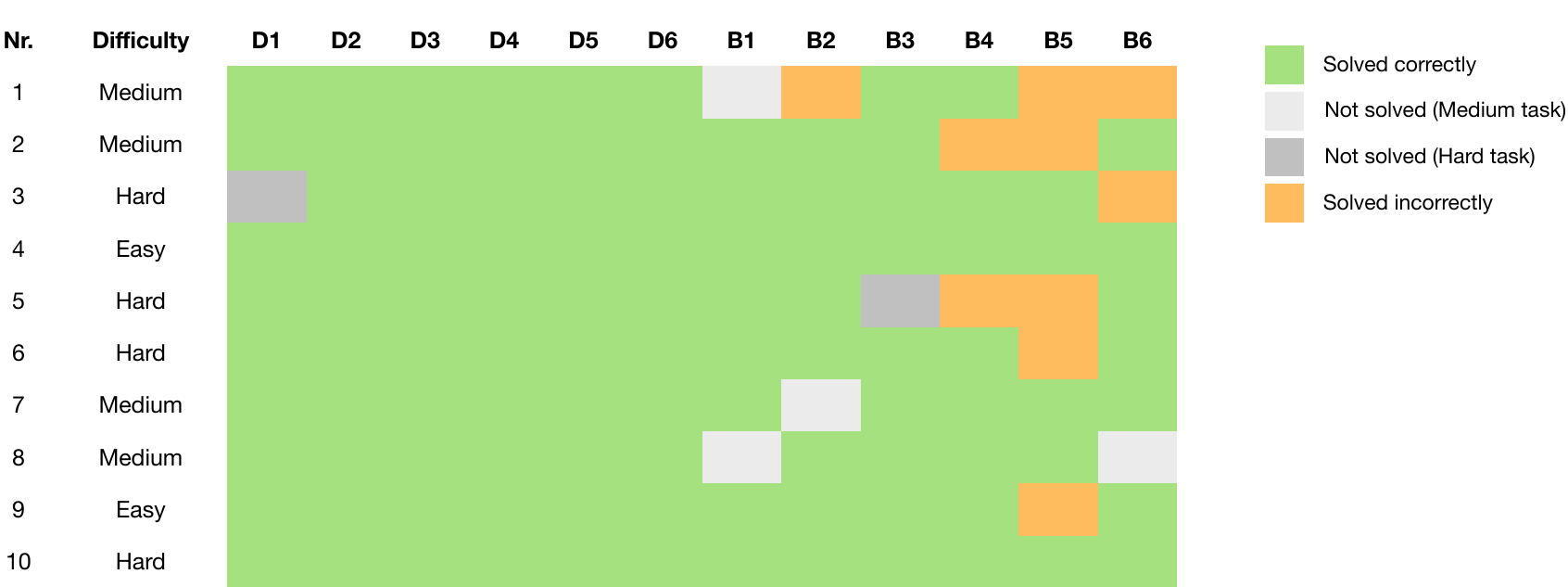}
    \caption{\textbf{Comparing the results between \sysName and the baseline condition.}
    \textbf{Overall results for each task for participants from \sysName condition (\DianaParticipant{1}-\DianaParticipant{6}) and from the baseline condition (\BaselineParticipant{1} - \BaselineParticipant{6}). Results indicate that the participants from the baseline condition performed worse than the participants from the \sysName condition.}}
    \label{fig:comparison-results}
    \Description{This table compares task outcomes for participants D1–D6 and B1–B6 across ten tasks, with icons indicating whether each task was solved correctly, solved incorrectly, or not solved. The pattern across the table shows that Hard tasks tend to yield the fewest correct solutions and the highest rates of failure or incomplete attempts. Medium tasks produce a mixture of correct and incorrect results, while Easy tasks are generally completed successfully by both groups. Although the difficulty gradient is consistent, the two participant groups differ in the types of tasks they handle well, and Diana participants perform better on interpretive tasks, while Baseline participants appear more successful on interaction-focused tasks.}
\end{figure}

Task 1 resulted in the highest number of incorrect responses among all tasks (Figure \ref{fig:comparison-results}).
Participants \BaselineParticipant{1}, \BaselineParticipant{2}, \BaselineParticipant{5}, and \BaselineParticipant{6} interacted extensively with the bar chart and other visuals in an attempt to locate the answer, even though the task only required a straightforward lookup and interpretation with no interaction.
\BaselineParticipant{3} and \BaselineParticipant{4} arrived at the correct answer but were unsure about its accuracy.
Task 2 also posed difficulties for \BaselineParticipant{4} and \BaselineParticipant{5}, who provided an incorrect answer after doing a series of manual calculations. 
Task 7, which required selecting a time period, was particularly demanding for \BaselineParticipant{2}.
She overlooked the time slicer entirely and focused only on the line chart showing monthly trends.
She then opened the visual in single-view mode using the Power BI built-in feature and attempted to apply dataset-level filters manually.
After struggling to inspect individual points in the line chart and then flipping to a table view, she eventually gave up.
Other participants solved the task, though \BaselineParticipant{6} expressed uncertainty even when her answer was correct.
Task 8 also led to a similar pattern of manual computation. \BaselineParticipant{1} said aloud \textit{``I guess I could add them. [...] Is there a way to select all, or is that just it?''}
Similar sentiments were expressed by \BaselineParticipant{4} and \BaselineParticipant{5}.
\BaselineParticipant{4} even wished for a similar feature as in Excel where points are added automatically: \textit{``just like what a[n] Excel table will do...
so that I don't have to calculate by myself.''}
This highlighted how unused or unfamiliar interaction techniques in Power BI led participants to revert to manual addition.
Overall, participants did not use keyboard-and-mouse interactions, including selecting data points to obtain aggregated values.

\subsubsection{Hard Tasks}

Similar to medium tasks, hard tasks were also computationally demanding, and participants were also uncertain about their answers.
Participants frequently questioned their answers (\BaselineParticipant{3}, \BaselineParticipant{5}, and \BaselineParticipant{6}) and, at times, provided incorrect responses or were unable to complete the tasks.
Each participant approached Task 3 differently.
\BaselineParticipant{1} discovered and used the drill-down interaction on the map visual to solve the task, whereas \BaselineParticipant{2}, \BaselineParticipant{3}, and \BaselineParticipant{4} relied on manual calculations rather than using the available interactivity.
Task 5, which also required drill-down interaction, was solved by only half of the participants.
\BaselineParticipant{3} and \BaselineParticipant{4} located the drill-down feature but were unable to use it effectively to reach the correct answer.
\BaselineParticipant{5} was unable to solve the task altogether; she waited for additional explanations on hover and appeared to expect the system to surface more guidance automatically.
\BaselineParticipant{6} often consulted the user guide to find answers and to see which visualizations were explained during the task-solving phase.

\subsection{Comparison Between \sysName and Static Dashboard Guide}

To better understand the effect of the onboarding method, we compared task performance (as shown in Figure \ref{fig:comparison-results}), answer correctness, and participants' overall experience using \sysName with those who relied solely on the static dashboard guide.

\subsubsection{Task Accuracy and Completeness}

Participants who used \sysName performed better overall than those relying solely on the user guide.
As task difficulty and cardinality increased, baseline condition participants struggled both to interpret the dashboard and to arrive at correct answers.
We distinguish between tasks participants could not solve or skipped (completeness) and those answered incorrectly (accuracy).
In both study groups, at least one task could not be solved by a participant.
In the \sysName group, this occurred only once: \DianaParticipant{1} was unable to solve Task 3.
In contrast, in the baseline condition group, five tasks were left unsolved: one each by \BaselineParticipant{2}, \BaselineParticipant{3}, and \BaselineParticipant{6}, and two by \BaselineParticipant{1}.
For example, \BaselineParticipant{6} could not complete Task 8 because she did not realize that, in her earlier interactions, she had applied filters, which had caused the map to hide relevant data points.

Incorrect answers were also more prevalent in the static dashboard guide group.
Across participants \BaselineParticipant{1}–\BaselineParticipant{6}, a total of ten tasks were answered incorrectly: five by \BaselineParticipant{5}, two each by \BaselineParticipant{4} and \BaselineParticipant{6}, and one by \BaselineParticipant{2}.
Despite differences in visualization and dashboard expertise, every participant in the guide group had at least one task that they either could not solve or answered incorrectly.

\subsubsection{Feeling of Uncertainty}

Although confidence was not an explicit study measure, we observed a notable contrast between the two groups in how they expressed uncertainty while solving tasks.
Participants using \sysName occasionally sought confirmation of their answers (e.g., \DianaParticipant{2} and \DianaParticipant{3}) and were successful in receiving it via one or more modalities.

In contrast, participants relying on the baseline condition expressed persistent uncertainty throughout the study.
They frequently used phrases such as \textit{``not sure''} or \textit{``not right, but...''} while reasoning aloud.
For instance, during Task 6, despite eventually providing the correct answer, \BaselineParticipant{6} remarked, \textit{``Okay, it’s definitely not right.
I don't think so.
But if this is what I have here...''}
Similar hesitations were observed for \BaselineParticipant{2}, \BaselineParticipant{3}, \BaselineParticipant{4}, \BaselineParticipant{5}, and \BaselineParticipant{6}, with each of them voicing uncertainty at least once during the session.

\subsubsection{Underexplored Interactivity and Reliance on Manual Calculations}

Participants using \sysName learned a range of interaction techniques through onboarding support and demonstrated a learning effect by Task 9.
In contrast, participants in the baseline group struggled with basic and intermediate interactivity throughout the study.
All participants in the baseline condition resorted to manually adding data-point values often performing arithmetic on paper because they did not understand or use the available keyboard-based interactions.
Microsoft Power BI dashboards require a combination of mouse and keyboard interactions (e.g., multi-selecting values, drilling down, removing filters), yet these participants relied almost entirely on mouse hovering or visual inspection.

\subsection{Reflections on \sysName}

In the post-task interview, participants (\DianaParticipant{1}-\DianaParticipant{6}) reflected on three main themes: their choice of modality, their experience with the onboarding, and the comparison to previous onboarding experiences.

\subsubsection{Preferred Modalities}

As participants conducted the user study using \sysName, their preferred modality became clear.
Despite varying levels of prior experience, all participants experimented with speech interaction. 
Even those with no prior exposure (\DianaParticipant{1}, \DianaParticipant{2}, \DianaParticipant{6}) adopted it during the study, with \DianaParticipant{2} initially hesitant but later describing it as \textit{``comfortable, almost not robotish.''}
Voice was valued for its speed and natural phrasing, while chat was preferred for step-by-step guidance (\DianaParticipant{1} and \DianaParticipant{5}).
Mouse interactions, such as lasso selection, were consistently used to define a focus before turning to another modality (\DianaParticipant{3} and \DianaParticipant{4}).

Participants also found multiple modalities useful, as they allowed them to adapt depending on time, task, or personal style.
\DianaParticipant{1} summarized this flexibility: \textit{``If you are in a hurry, you can just talk to it, and it goes faster.
If you have time to sit down and think, then you can use the chat and the radial menu...
But I would probably use writing at most, because I’m old-fashioned.''}

Across modalities, the visual highlight feature was universally appreciated to orient user attention.

\subsubsection{Strengths of Voice-based Interaction and Visual Highlights Feature}

Speech interaction was well received, even among those without prior experience.
\DianaParticipant{2} remarked that \textit{``it was very comfortable voice, I could hear that it was mechanical, but it was not that mechanical.''}
He also appreciated the real-time response, and when paired with the visual highlighting feedback, it helped him see what \sysName was saying in response to his question (\DGFour).
As \DianaParticipant{2} was an expert, he could verify the correctness of the answer provided by \sysName.
As the answer was phrased naturally and accurately, he said \textit{``and when it did that [gave him the right steps to the answer] it had my trust, that I could actually ask the voice interaction a question.''}
\DianaParticipant{3} described the voice modality and the response from \sysName as informative and that he \textit{``really [had] fun interacting with her,''} while \DianaParticipant{4} appreciated that speech \textit{``doesn’t stop when you are doing something else on the dashboard.''}
The mouse modality, in the form of lasso selection, was also valued when combined with voice, as the combination helped participants anchor their queries to specific visuals.

\subsubsection{Underused Modalities and Features}

The radial menu was the least used feature.
While \DianaParticipant{2} found it helpful for confirmation, most others described it as unintuitive, overly detailed, or unfamiliar:
\textit{``I didn’t know what specifically to use it for [...] the surface things were good enough''} (\DianaParticipant{4}).
The user guide was largely ignored, with \DianaParticipant{1} and \DianaParticipant{3} admitting they had forgotten it existed.
\DianaParticipant{6} said she would have preferred to read it if time allowed.
This indicates that static documentation can be easily overshadowed by interactive support.
Although useful for step-by-step guidance, the chat via keyboard was sometimes described as slow or effortful (\textit{``chatting took too long to write compared to just speaking,''} \DianaParticipant{2}).

\subsubsection{Comparison to Prior Onboarding.}

Compared to step-by-step onboarding methods, \sysName was described as more flexible and empowering.
\DianaParticipant{4} compared it to video game tutorials where you are forced to proceed linearly:
\textit{``You have to explore one step at a time... Here you have the freedom to onboard yourself the way you want to.''}
\DianaParticipant{1} echoed this, preferring to ask for help when needed rather than \textit{``click all over''} through a fixed tutorial (\DGFive).
For \DianaParticipant{2}, who regularly onboards colleagues through videos and one-on-one sessions, \sysName represented a potential time-saver.
Although he values personal interaction, he saw the system as a \textit{``virtual project finance controller''} (\DGThree) that could reduce his workload.

%% TODO: How does the experience relate to the user level.
\section{Discussion}
We now discuss the implications of our findings in the context of dashboard onboarding and the use of large language models for onboarding.
\subsection{Speech Interaction as the Human Onboarder Assistant}

In the main study, participants frequently preferred speech over mouse or keyboard-based interactions for onboarding help and follow-up questions.
This result matches the findings of León et al.~\cite{molina24} on modality preferences for exploration tasks.
This is also in line with the findings from Zhan et al.~\cite{healthcareva} who studied trust in healthcare voice assistants and suggested that perceived usefulness of a system is a strong indicator of the trust a user has in the system. 

Rather than replacing human onboarders, \sysName assists and augments their capabilities by handling routine questions such as \textit{how do I...?} that are time-consuming to document and maintain. 
In the context of onboarding, the result also aligns with one of our initial goals of providing human-like support ({\DGThree}).
The findings suggest that participants appreciated real-time responses and being less dependent on asking a human expert for help.

In our formative study, we also came across a surprising result: 
the Realtime API~\cite{realtime} detected that the participant was not a native English speaker and switched to another language that the person understood.
While unexpected, the person was positively surprised by the detection. 
Furthermore, other participants wished for natural language interaction in their native language.
While this is an option with current large language models, it may go beyond the capabilities of many human onboarders. 

\subsection{Multimodal Interaction for Onboarding}

The different preferences and interaction patterns we found suggest that multimodality is a suitable approach for onboarding.
While some participants had a clear preference for one modality, others, such as keyboard-based interaction for chat or voice-based interaction, others preferred combining two or more. This was visible when the tasks were difficult.

The less use of the contextual menu could also suggest that sophisticated visual encodings for help are not necessary for onboarding, as they were perceived as advanced or quite detailed. But, they would be extremely helpful at later stages after users have been onboarded, to allow them to access more fine-grained information, as echoed by \DianaParticipant{1}.

\subsection{Limited Adoption of Radial Menu}

Even though the radial menu was designed to provide structured and faster guidance compared to other LLM-powered onboarding options, such as chat and speech interaction, it was the least used feature overall.
Participants reported several reasons for this, such as unintuitive (\DianaParticipant{3}), too much information (\DianaParticipant{2}, \DianaParticipant{4}), and generally time-consuming to find answers (\DianaParticipant{1}).
In addition, the unfamiliar design (\DianaParticipant{4}) and the presence of technical terms may have hindered adoption.
Unlike the LLM-powered onboarding features, which provided help tailored to users' needs, the radial menu required participants to actively explore and discover relevant information, which could also explain its low usage. We plan to improve the radial menu design based on feedback from the user study.

\subsection{Using Large Language Models for Onboarding}

While the natural interaction via voice and text was successful, it required a substantial effort at the technical level to minimize the number of hallucinations. 
The information sent to the LLMs was carefully designed based on available REST API information and heuristics to ensure that information from the REST APIs is not wildly extrapolated.
The prompts were fine-tuned several times throughout the formative study.
We were able to significantly reduce hallucinations in both chat- and voice-based interactions, but participants still encountered issues when they highlighted a specific section of the dashboard and asked for guidance via natural language. 

Additionally, because \sysName has no access to the data, it can be used in any domain with ease, as it only provides information about the high-level aspects of the visualization.
This was an intentional design choice motivated by privacy and security concerns, and it also prevented the LLM from giving direct answers, which participants attempted to obtain on several occasions.
It can infer trends and drivers, however, it cannot give detailed data-level information that is usually critical in a business context.
With controlled access to data values and their explanations, the system could provide richer, more contextualized guidance during dashboard exploration.

\subsection{Limitations}

We examine the limitations of \sysName\ with respect to its design, generalizability, deployment, and practical considerations, including cost. We also outline the limitations of our user study.

\subsubsection{Design Limitations}

The design limitations of \sysName became evident during the user study.
The low usage of the radial menu calls for a redesign, including simplifying its hierarchy or reducing the amount of information shown at once.
Beyond simplification, different forms of contextual or on-visual menus may also need to be explored.
Additionally, sometimes the visual highlights were not synchronized with the voice output, causing confusion.

\subsubsection{Generalizability}

\inserted{\sysName supports Microsoft Power BI dashboards with default visuals. Many of its core design concepts—such as the Vue-based frontend components, multimodal interaction features, and an LLM-backed backend with structured prompting—could be generalized to other dashboard platforms and custom visuals. The design concepts may generalize for custom visuals and new dashboard platforms, however technical implementation would require some engineering effort, such as:}

\deleted {We developed \mbox{\sysName} specifically for Microsoft Power BI, and it currently works with any dashboard with default visuals.
It can also support custom visuals, but only if developers supply the necessary metadata, as Power BI does not expose this information by default.}

\deleted{The frontend is implemented as a set of Vue-based components.
Therefore, many of the multimodal interaction features could be reused in other dashboarding environments.
The backend has the LLM integration and the prompting structure, which are also conceptually transferable.}

\deleted{While the design concepts may generalize, the technical implementation would require some adaptation to function in other dashboard systems.}

{\deleted{Cross-Platform Deployment}}

\deleted{Integrating \mbox{\sysName} to non-Power BI dashboards may require some engineering effort, such as:}

\begin{itemize}
    \item Metadata extraction: For new platforms, visual metadata would either need to be structured according to the current implementation or the implementation adapted to the new platform’s metadata format.
    \item Interaction support: Power BI uses keyboard and mouse interactions, which may differ or be absent on other platforms. Therefore, prompts and interaction handling would need to be adjusted accordingly.
\end{itemize}.

Cross-platform deployment could enable broader applicability of \sysName, but we have not yet empirically validated it beyond Power BI.
Future work could explore prototyping on alternative dashboard platforms to confirm feasibility and usability.

\subsubsection{Cost Considerations}

We also relied on OpenAI APIs, including the Realtime API, which incurs usage-based costs.
The Realtime API \cite{realtime} can become expensive depending on interaction volume.
As AI models continue to improve, these costs may decrease, and more affordable alternatives may emerge.
However, the current API pricing remains a practical concern for the widespread deployment and generalizability of \sysName.

\subsubsection{User Study Limitations}

The user study was conducted on a single dashboard to avoid bias, ensuring consistent difficulty and controlled cardinality (i.e., the number of views a task required).
However, this choice limits the generalizability of the findings to that specific dashboard.
Additionally, the two study groups differed slightly in their prior expertise, which may have influenced performance differences. \inserted{Specifically, participants in the \mbox{\sysName} condition reported slightly greater prior experience with dashboards compared to those in the static dashboard guide group.}

Future work could investigate learning improvement and time efficiency more systematically.
This would require a longitudinal study in which \sysName is deployed over an extended period and compared against multiple baseline onboarding methods, such as static guides, video tutorials, or step-by-step tutorials, to provide quantitative evidence of its effectiveness.

\section{Conclusion}

We designed and developed \sysName, a multimodal dashboard assistant that combines voice, text, in-situ visual highlights, and contextual help to support users during dashboard onboarding. 
In an exploratory user study, we found that participants using \mbox{\sysName} achieved higher task accuracy and completeness than those using the static dashboard guide. Participants in the \sysName condition used all interaction modalities and preferred voice-based interaction over keyboard- and mouse-based interaction. 
Our findings suggest that LLMs are useful for onboarding tasks when they are paired with visual affordances. It also encouraged users to ask for help without compromising their autonomy.
We suggest that onboarding systems should be multimodal by design and offer a low-friction path for users to onboard themselves on dashboards.

\begin{acks}
    This work was supported partly by Villum Investigator grant VL-54492 by Villum Fonden, the Austrian Science Fund (FWF DFH 23–N), and the Austrian Research Promotion Agency (FFG 911655).
    The VRVis GmbH is funded by BMIMI, BMWET, Tyrol, Vorarlberg and Vienna Business Agency in the scope of COMET - Competence Centers for Excellent Technologies (911654) which is managed by FFG.
    Any opinions, findings, and conclusions expressed in this material are those of the authors and do not necessarily reflect the views of the funding agency.
\end{acks}

\bibliographystyle{ACM-Reference-Format}
\bibliography{main}

@String{addrACM           = {New York, NY, USA}}

@String{addrIEEECS        = {Los Alamitos, CA, USA}}

@String{jourCGF           = {Computer Graphics Forum}}

@String{jourCompGraphAppl = {{IEEE} Computer Graphics and Applications}}

@String{jourIVS           = {Information Visualization}}

@String{jourTVCG          = {{{IEEE} Transactions on Visualization and Computer Graphics}}}

@String{jourVI            = {{Visual Informatics}}}

@String{procCHI           = {Proceedings of the {ACM} Conference on Human Factors in Computing Systems}}

@String{procITS           = {Proceedings of the {ACM} Conference on Interactive Tabletops and Surfaces}}

@String{procACMHCI     = {Proceedings of the {ACM} Conference on Human--Computer Interaction}}

@String{procVisCom        = {Proceedings of the Visualization for Communication Workshop at IEEE VIS}}

@String{procAVI          = {Proceedings of the {ACM} Conference on Advanced Visual Interfaces}}

@String{pubACM            = {{ACM}}}

@String{pubIEEECS           = {{IEEE Computer Society}}}

@article{DBLP:journals/tvcg/ChunduryPRTLE22,
  author       = {Pramod Chundury and
                  Biswaksen Patnaik and
                  Yasmin Reyazuddin and
                  Christine Tang and
                  Jonathan Lazar and
                  Niklas Elmqvist},
  title        = {Towards Understanding Sensory Substitution for Accessible Visualization: An Interview Study},
  journal      = jourTVCG,
  volume       = {28},
  number       = {1},
  pages        = {1084--1094},
  year         = {2022},
  doi          = {10.1109/TVCG.2021.3114829},
}

@article{DBLP:journals/corr/abs-2303-02927,
  author       = {Victor Dibia},
  title        = {{LIDA:} {A} Tool for Automatic Generation of Grammar-Agnostic Visualizations and Infographics using Large Language Models},
  journal      = {CoRR},
  volume       = {abs/2303.02927},
  year         = {2023},
  doi          = {10.48550/ARXIV.2303.02927},
  eprinttype   = {arXiv},
  eprint       = {2303.02927},
  numpages     = {14},
}

@article{DBLP:journals/tvcg/DengWQW23,
  author       = {Dazhen Deng and
                  Aoyu Wu and
                  Huamin Qu and
                  Yingcai Wu},
  title        = {{DashBot}: Insight-Driven Dashboard Generation Based on Deep Reinforcement
                  Learning},
  journal      = jourTVCG,
  volume       = {29},
  number       = {1},
  pages        = {690--700},
  year         = {2023},
  doi          = {10.1109/TVCG.2022.3209468},
}

@article{DBLP:journals/tvcg/ZhaoWXZGTZC25,
  author       = {Yuheng Zhao and
                  Junjie Wang and
                  Linbing Xiang and
                  Xiaowen Zhang and
                  Zifei Guo and
                  Cagatay Turkay and
                  Yu Zhang and
                  Siming Chen},
  title        = {{LightVA}: Lightweight Visual Analytics With {LLM} Agent-Based Task
                  Planning and Execution},
  journal      = jourTVCG,
  volume       = {31},
  number       = {9},
  pages        = {6162--6177},
  year         = {2025},
  doi          = {10.1109/TVCG.2024.3496112},
}

@article{DBLP:journals/tvcg/WuWZHZQZ22,
  author       = {Aoyu Wu and
                  Yun Wang and
                  Mengyu Zhou and
                  Xinyi He and
                  Haidong Zhang and
                  Huamin Qu and
                  Dongmei Zhang},
  title        = {{MultiVision}: Designing Analytical Dashboards with Deep Learning Based
                  Recommendation},
  journal      = jourTVCG,
  volume       = {28},
  number       = {1},
  pages        = {162--172},
  year         = {2022},
  doi          = {10.1109/TVCG.2021.3114826},
}

@inproceedings{DBLP:conf/sigmod/KeyHPA12,
  author       = {Alicia Key and
                  Bill Howe and
                  Daniel Perry and
                  Cecilia R. Aragon},
  title        = {VizDeck: self-organizing dashboards for visual analytics},
  booktitle    = {Proceedings of the {ACM} Conference on Management of Data},
  pages        = {681--684},
  publisher    = pubACM,
  address      = addrACM,
  year         = {2012},
  doi          = {10.1145/2213836.2213931},
}

@article{DBLP:journals/tvcg/ElshehalyRBMAGR21,
  author       = {Mai Elshehaly and
                  Rebecca Randell and
                  Matthew Brehmer and
                  Lynn McVey and
                  Natasha Alvarado and
                  Chris P. Gale and
                  Roy A. Ruddle},
  title        = {{QualDash}: Adaptable Generation of Visualisation Dashboards for Healthcare
                  Quality Improvement},
  journal      = jourTVCG,
  volume       = {27},
  number       = {2},
  pages        = {689--699},
  year         = {2021},
  doi          = {10.1109/TVCG.2020.3030424},
}

@article{DBLP:journals/corr/abs-2208-03175,
  author       = {Aditeya Pandey and
                  Arjun Srinivasan and
                  Vidya Setlur},
  title        = {{MEDLEY:} Intent-based Recommendations to Support Dashboard Composition},
  journal      = {CoRR},
  volume       = {abs/2208.03175},
  year         = {2022},
  doi          = {10.48550/ARXIV.2208.03175},
  eprinttype   = {arXiv},
  eprint       = {2208.03175},
  numpages     = {11},
}

@article{DBLP:journals/tvcg/JakobsenHKH13,
  author       = {Mikkel R. Jakobsen and
                  Yonas Sahlemariam Haile and
                  S{\o}ren Knudsen and
                  Kasper Hornb{\ae}k},
  title        = {Information Visualization and Proxemics: Design Opportunities and Empirical Findings},
  journal      = jourTVCG,
  volume       = {19},
  number       = {12},
  pages        = {2386--2395},
  year         = {2013},
  doi          = {10.1109/TVCG.2013.166},
}

@article{DBLP:journals/tvcg/SrinivasanS18,
  author       = {Arjun Srinivasan and
                  John T. Stasko},
  title        = {Orko: Facilitating Multimodal Interaction for Visual Exploration and Analysis of Networks},
  journal      = jourTVCG,
  volume       = {24},
  number       = {1},
  pages        = {511--521},
  year         = {2018},
  doi          = {10.1109/TVCG.2017.2745219},
}

@inproceedings{DBLP:conf/ieeevast/BadamAEI16,
  author       = {Sriram Karthik Badam and
                  Fereshteh Amini and
                  Niklas Elmqvist and
                  Pourang Irani},
  title        = {Supporting visual exploration for multiple users in large display environments},
  booktitle    = {Proceedings of the {IEEE} Conference on Visual Analytics Science and Technology},
  pages        = {1--10},
  publisher    = pubIEEECS,
  address      = addrIEEECS,
  year         = {2016},
  doi          = {10.1109/VAST.2016.7883506},
}

@inproceedings{DBLP:conf/tabletop/BaurLC12,
  author       = {Dominikus Baur and
                  Bongshin Lee and
                  Sheelagh Carpendale},
  title        = {{TouchWave}: kinetic multi-touch manipulation for hierarchical stacked
                  graphs},
  booktitle    = procITS,
  pages        = {255--264},
  publisher    = pubACM,
  address      = addrACM,
  year         = {2012},
  doi          = {10.1145/2396636.2396675},
}

@inproceedings{DBLP:conf/avi/ThompsonSS18,
  author       = {John Thompson and
                  Arjun Srinivasan and
                  John T. Stasko},
  title        = {Tangraphe: interactive exploration of network visualizations using single hand, multi-touch gestures},
  booktitle    = procAVI,
  pages        = {43:1--43:5},
  publisher    = pubACM,
  address      = addrACM,
  year         = {2018},
  doi          = {10.1145/3206505.3206519},
}

@inproceedings{DBLP:conf/vissym/SrinivasanS17,
  author       = {Arjun Srinivasan and
                  John T. Stasko},
  title        = {Natural Language Interfaces for Data Analysis with Visualization: Considering What Has and Could Be Asked},
  booktitle    = {Short Paper Proceedings of the Eurographics Conference on Visualization},
  pages        = {55--59},
  publisher    = {Eurographics Association},
  address      = {Eindhoven, The Netherlands},
  year         = {2017},
  doi          = {10.2312/EUROVISSHORT.20171133},
}

@inproceedings{DBLP:conf/ozchi/NielsenEG16,
  author       = {Matthias Nielsen and
                  Niklas Elmqvist and
                  Kaj Gr{\o}nb{\ae}k},
  title        = {Scribble query: fluid touch brushing for multivariate data visualization},
  booktitle    = {Proceedings of the Australian Conference on Computer-Human Interaction},
  pages        = {381--390},
  publisher    = pubACM,
  address      = addrACM,
  year         = {2016},
  doi          = {10.1145/3010915.3010951},
}

@inproceedings{c_visualization_2019,
    title = {Visualization Onboarding: Learning How to Read and Use Visualizations},
    author = {Christina Stoiber and Florian Grassinger and Margit Pohl and Holger Stitz and Marc Streit and Wolfgang Aigner},
    booktitle = procVisCom,
    doi = {10.31219/osf.io/c38ab},
    year = {2019}, 
    numpages = {6},
    publisher = pubIEEECS,
    address = addrIEEECS,
}

@article{DBLP:journals/tvcg/DhanoaHFEGS25,
  author       = {Vaishali Dhanoa and
                  Andreas P. Hinterreiter and
                  Vanessa Fediuk and
                  Niklas Elmqvist and
                  Eduard Gr{\"{o}}ller and
                  Marc Streit},
  title        = {{D-Tour}: Semi-Automatic Generation of Interactive Guided Tours for
                  Visualization Dashboard Onboarding},
  journal      = jourTVCG,
  volume       = {31},
  number       = {1},
  pages        = {721--731},
  year         = {2025},
  doi          = {10.1109/TVCG.2024.3456347},
}

@inproceedings{DBLP:conf/chi/ElmqvistTT08,
  author       = {Niklas Elmqvist and
                  Mihail Eduard Tudoreanu and
                  Philippas Tsigas},
  title        = {Evaluating motion constraints for {3D} wayfinding in immersive and desktop virtual environments},
  booktitle    = procCHI,
  pages        = {1769--1778},
  publisher    = pubACM,
  address      = addrACM,
  year         = {2008},
  doi          = {10.1145/1357054.1357330},
}

@article{a_what_2018,
  author       = {Alper Sarikaya and
                  Michael Correll and
                  Lyn Bartram and
                  Melanie Tory and
                  Danyel Fisher},
  title        = {What Do We Talk About When We Talk About Dashboards?},
  journal      = jourTVCG,
  volume       = {25},
  number       = {1},
  pages        = {682--692},
  year         = {2019},
  doi          = {10.1109/TVCG.2018.2864903},
}

@inproceedings{m_annotating_2012,
	title = {Annotating {BI} visualization dashboards: needs and challenges},
	doi = {10.1145/2207676.2208288},
	booktitle = procCHI,
	author = {Micheline Elias and Anastasia Bezerianos},
	year = {2012},
	pages = {1641--1650},
        publisher = pubACM,
        address = addrACM,
}

@article{ceneda_characterizing_2017,
	title = {Characterizing Guidance in Visual Analytics},
	volume = {23},
	doi = {10.1109/TVCG.2016.2598468},
	number = {1},
	journal = jourTVCG,
	author = {Ceneda, Davide and Gschwandtner, Theresia and May, Thorsten and Miksch, Silvia and Schulz, Hans-J{\"o}rg and Streit, Marc and Tominski, Christian},
	month = jan,
	year = {2017},
	pages = {111--120},
}

@article{collins_guidance_2018,
	title = {Guidance in the human–machine analytics process},
	volume = {2},
	doi = {10.1016/j.visinf.2018.09.003},
	number = {3},
	journal = jourVI,
	author = {Collins, Christopher and Andrienko, Natalia and Schreck, Tobias and Yang, Jing and Choo, Jaegul and Engelke, Ulrich and Jena, Amit and Dwyer, Tim},
	month = sep,
	year = {2018},
	pages = {166--180},
}

@article{walchshofer_transitioning_2023,
  title     = {Transitioning to a Commercial Dashboarding System: Socio-technical Observations and Opportunities},
  doi       = {10.1109/TVCG.2023.3326525},
  journal   = jourTVCG,
  author    = {Walchshofer, Conny and Dhanoa, Vaishal and Streit, Marc and Meyer, Miriah},
  year      = {2023},
  volume    = {30},
  number    = {1},
  pages     = {381--391},
}

@article{stoiber_comparative_2022,
	title = {Comparative evaluations of visualization onboarding methods},
	volume = {6},
	doi = {10.1016/j.visinf.2022.07.001},
	number = {4},
	journal = jourVI,
	author = {Stoiber, Christina and Walchshofer, Conny and Pohl, Margit and Potzmann, Benjamin and Grassinger, Florian and Stitz, Holger and Streit, Marc and Aigner, Wolfgang},
	month = dec,
	year = {2022},
	pages = {34--50},
}

@article{tory_finding_2022,
	title = {Finding {Their} {Data} {Voice}: {Practices} and {Challenges} of {Dashboard} {Users}},
	doi = {10.1109/MCG.2021.3136545},	
	urldate = {2022-11-09},
	journal = jourCompGraphAppl,
	author = {Tory, Melanie and Bartram, Lyn and Fiore-Gartland, Brittany and Crisan, Anamaria},
	year = {2021},
	pages = {22--36},
        volume = {43},
        number = {1},
}

@book{few_information_2006,	
	title = {Information Dashboard Design: The Effective Visual Communication of Data},
	publisher = {O'Reilly Media},
	author = {Few, Stephen},
	year = {2006},
	address = {Sebastopol, CA, USA},
}

@article{DBLP:journals/ivs/ChunduryYCMSE23,
  author       = {Pramod Chundury and Mehmet Adil Yal{\c{c}}in and Jonathan Crabtree and Anup Mahurkar and Lisa M. Shulman and Niklas Elmqvist},
  title        = {Contextual in situ help for visual data interfaces},
  journal      = jourIVS,
  volume       = {22},
  number       = {1},
  pages        = {69--84},
  year         = {2023},
  doi          = {10.1177/14738716221120064},
}

@article{v_process_2022,
	title = {A Process Model for Dashboard Onboarding},
	doi = {10.1111/cgf.14558},
	journal = jourCGF,
        volume = {41},
        pages = {501--513},
	author = {Vaishali Dhanoa and C. Walchshofer and Andreas Hinterreiter and Holger Stitz and Eduard Groeller and Marc Streit},
	year = {2022},
}

@ARTICLE{bach23dashboardpatterns_tvcg,
  author    = {Bach, Benjamin and Freeman, Euan and Abdul-Rahman, Alfie and Turkay, Cagatay and Khan, Saiful and Fan, Yulei and Chen, Min},
  journal   = jourTVCG,
  title     = {Dashboard Design Patterns}, 
  year      = {2023},
  volume    = {29},
  number    = {1},
  pages     = {342--352},
  doi       = {10.1109/TVCG.2022.3209448},
}

@article{DBLP:journals/tvcg/SarikayaCBTF19,
  author       = {Alper Sarikaya and
                  Michael Correll and
                  Lyn Bartram and
                  Melanie Tory and
                  Danyel Fisher},
  title        = {What Do We Talk About When We Talk About Dashboards?},
  journal      = jourTVCG,
  volume       = {25},
  number       = {1},
  pages        = {682--692},
  year         = {2019},
  doi          = {10.1109/TVCG.2018.2864903},
}

@InProceedings{Badam2017,
  author    = {Sriram Karthik Badam and Arjun Srinivasan and Niklas Elmqvist and John Stasko},
  title     = {Affordances of Input Modalities for Visual Data Exploration in Immersive Environments},
  booktitle = {Proceedings of the IEEE VIS Immersive Analytics Workshop},
  publisher = pubIEEECS,
  address   = addrIEEECS,
  year      = {2017},
  numpages  = {5},
  url       = {https://karthikbadam.github.io/files/mixing-modalities.pdf},
}

@ARTICLE{lee12beyond,  
  author={Lee, Bongshin and Isenberg, Petra and Riche, Nathalie Henry and Carpendale, Sheelagh},  
  journal=jourTVCG,  
  title={Beyond Mouse and Keyboard: Expanding Design Considerations for Information Visualization Interactions},
  year={2012},  
  volume={18},  
  number={12},  
  pages={2689--2698},  
  doi={10.1109/TVCG.2012.204},  
}

@ARTICLE{saktheeswaran20,
author={Saktheeswaran, Ayshwarya and Srinivasan, Arjun and Stasko, John},
journal=jourTVCG,
title={Touch? Speech? or Touch and Speech? {I}nvestigating Multimodal Interaction for Visual Network Exploration and Analysis},
year={2020},
volume={26},
number={6},
pages={2168--2179},
doi={10.1109/TVCG.2020.2970512},
}

@inbook{srinivasan20inchorus,
author = {Srinivasan, Arjun and Lee, Bongshin and Henry Riche, Nathalie and Drucker, Steven M. and Hinckley, Ken},
title = {{InChorus}: Designing Consistent Multimodal Interactions for Data Visualization on Tablet Devices},
year = {2020},
isbn = {9781450367080},
publisher = pubACM,
address = addrACM,
doi = {10.1145/3313831.3376782},
url = {https://doi.org/10.1145/3313831.3376782},
booktitle = procCHI,
pages = {1--13},
numpages = {13}
}

@inproceedings{srinivasan2023bolt,
  author       = {Arjun Srinivasan and
                  Vidya Setlur},
  title        = {{BOLT:} {A} Natural Language Interface for Dashboard Authoring},
  booktitle    = {Short Paper Proceedings of the Eurographics Conference on Visualization},
  pages        = {7--11},
  publisher    = {Eurographics Association},
  address      = {Eindhoven, The Netherlands},
  year         = {2023},
  doi          = {10.2312/EVS.20231035},
}

@article{DBLP:journals/ivs/ElmqvistMJCRJ11,
  author       = {Niklas Elmqvist and
                  Andrew Vande Moere and
                  Hans{-}Christian Jetter and
                  Daniel Cernea and
                  Harald Reiterer and
                  T. J. Jankun{-}Kelly},
  title        = {Fluid interaction for information visualization},
  journal      = jourIVS,
  volume       = {10},
  number       = {4},
  pages        = {327--340},
  year         = {2011},
  doi          = {10.1177/1473871611413180},
}

@ARTICLE{lin24,
  author={Lin, Yanna and Li, Haotian and Wu, Aoyu and Wang, Yong and Qu, Huamin},
  journal=jourTVCG, 
  title={D{M}iner: Dashboard Design Mining and Recommendation}, 
  year={2024},
  volume={30},
  number={7},
  pages={4108-4121},
  doi={10.1109/TVCG.2023.3251344}
}

@ARTICLE {molina24,
author = {Molina León, Gabriela and Isenberg, Petra and Breiter, Andreas},
journal = jourTVCG,
title = {Eliciting Multimodal and Collaborative Interactions for Data Exploration on Large Vertical Displays},
year = {2024},
volume = {30},
number = {2},
pages = {1624--1637},
doi = {10.1109/TVCG.2023.3323150},
}

@inproceedings{lisnic25plume,
  author = {Lisnic, Maxim and Setlur, Vidya and Sultanum, Nicole},
  title = {Plume: Scaffolding Text Composition in Dashboards},
  year = {2025},
  publisher = pubACM,
  address = addrACM,
  doi = {10.1145/3706598.3713580},
  booktitle = procCHI,
  pages = {1085:1--1085:18},
}

@misc{wen25prompt,
  title={Exploring Multimodal Prompt for Visualization Authoring with Large Language Models},
  author={Zhen Wen and Luoxuan Weng and Yinghao Tang and Runjin Zhang and Yuxin Liu and Bo Pan and Minfeng Zhu and Wei Chen},
  year={2025},
  eprint={2504.13700},
  archivePrefix={arXiv},
  primaryClass={cs.HC},
  url={https://arxiv.org/abs/2504.13700}, 
}

@article{hoque25dashguide,
  author       = {Md. Naimul Hoque and
                  Nicole Sultanum},
  title        = {{DashGuide}: Authoring Interactive Dashboard Tours for Guiding Dashboard Users},
  journal      = jourCGF,
  volume       = {44},
  number       = {3},
  year         = {2025},
  doi          = {10.1111/CGF.70107},
  articleno    = {e70107},
  numpages     = {12},
}

@article{chen25interchat,
  author       = {Juntong Chen and
                  Jiang Wu and
                  Jiajing Guo and
                  Vikram Mohanty and
                  Xueming Li and
                  Jorge Piazentin Ono and
                  Wenbin He and
                  Liu Ren and
                  Dongyu Liu},
  title        = {{InterChat}: Enhancing Generative Visual Analytics using Multimodal
                  Interactions},
  journal      = jourCGF,
  volume       = {44},
  number       = {3},
  year         = {2025},
  articleno    = {e70112},
  numpages     = {12},
  doi          = {10.1111/CGF.70112},
}

@misc{microsoft_restapi,
	title = {Using the {Power} {BI} {REST} {APIs}},
    author = {Kesem Sharabi and Mahir Diab and {laurent-mic} and Owen Duncan and Jian Dong and Sudeep Kumar},
	howpublished = {\url{https://learn.microsoft.com/en-us/rest/api/power-bi/}},
    note = {Last accessed on 2025/09/09},
    urldate = {2025/09/09},
	date = {2023-08-31},
    year = {2023},
}

@misc{microsoft_embedded,
	title = {Power {BI} {Embedded}},
    author = {{Microsoft Corporation}},
	howpublished = {\url{https://azure.microsoft.com/en-us/products/power-bi-embedded/}},
    note = {Last accessed on 2025/09/09},
    urldate = {2025/09/09},
	year = {2023},
}

@misc{vue,
	title = {Vue JS},
    author = {Evan You},
	howpublished = {\url{https://vuejs.org/}},
    note = {Last accessed on 2025/09/09},
    urldate = {2025/09/09},
	year = {2023},
}

@misc{realtime,
	title = {Realtime API},
    author = {{OpenAI}},
	howpublished = {\url{https://platform.openai.com/docs/guides/realtime}},
    note = {Last accessed on 2025/09/09},
    urldate = {2025/09/09},
	year = {2023},
}

@misc{gptmini,
	title = {GPT-4.1 mini},
    author = {{OpenAI}},
	howpublished = {\url{https://platform.openai.com/docs/models/gpt-4.1-mini}},
    note = {Last accessed on 2025/09/09},
    urldate = {2025/09/09},
	year = {2023},
}

@misc{flask,
title = {Flask},
    author = {{Pallets}},
	howpublished = {\url{https://flask.palletsprojects.com/en/stable/}},
    note = {Last accessed on 2025/09/09},
    urldate = {2025/09/09},
	year = {2023},
}

@misc{examplebi,
title = {Revenue Opportunities sample for Power BI},
    author = {{Microsoft Corporation}},
	howpublished = {\url{https://learn.microsoft.com/en-us/power-bi/create-reports/sample-revenue-opportunities/}},
    note = {Last accessed on 2025/09/09},
    urldate = {2025/09/09},
	year = {2023},
}

@article{healthcareva,
  author = {Zhan, Xiao and Abdi, Noura and Seymour, William and Such, Jose},
  title = {Healthcare Voice {AI} Assistants: Factors Influencing Trust and Intention to Use},
  year = {2024},
  publisher = pubACM,
  address = addrACM,
  volume = {8},
  number = {CSCW1},
  doi = {10.1145/3637339},
  journal = procACMHCI,
  month = apr,
  articleno = {62},
  numpages = {37},
}

@inproceedings{ko2004designing,
  title={Designing the whyline: a debugging interface for asking questions about program behavior},
  author={Ko, Amy J. and Myers, Brad A.},
  booktitle=procCHI,
  publisher = pubACM,
  address = addrACM,
  pages={151--158},
  doi = {10.1145/1842993.1843021},
 year={2004}
}

@inproceedings{menus,
  author = {Pourmemar, Majid and Poullis, Charalambos},
  title = {Visualizing and Interacting with Hierarchical Menus in Immersive Augmented Reality},
  year = {2019},
  publisher = pubACM,
  address = addrACM,
  doi = {10.1145/3359997.3365693},
  booktitle = {Proceedings of the ACM SIGGRAPH Conference on Virtual-Reality Continuum and Its Applications in Industry},
  articleno = {30},
  numpages = {9},
}

@inproceedings{10.1145/1842993.1843021,
  author = {Samp, Krystian and Decker, Stefan},
  title = {Supporting menu design with radial layouts},
  year = {2010},
  publisher = pubACM,
  address = addrACM,
  doi = {10.1145/1842993.1843021},
  booktitle = {Proceedings of the ACM Conference on Advanced Visual Interfaces},
  pages = {155–162},
  numpages = {8},
}
\end{document}